# L'effet de levier de trésorerie

La Théorie Financière attire notre attention sur les conséquences du poids des charges de structure dans l'activité d'une entreprise. *L'effet de levier d'exploitation* exprime le taux de variation du résultat provenant d'un certain taux de variation de la production vendue. L'entreprise qui a les coûts fixes les plus élevés pour un volume d'activité donné a le plus fort coefficient d'effet de levier, ce qui signifie que sa profitabilité est la plus sensible aux variations de la conjoncture. Dans le domaine de l'étude du *risque d'insolvabilité* de la firme, une adaptation du *concept d'élasticité* à l'analyse de la sensibilité de la trésorerie semble pertinente, dans la mesure où elle met en relief les exigences imposées par la nature de la combinaison productive. *L'effet de levier de trésorerie* porte un éclairage particulier sur les risques de défaillance liés au choix de la combinaison productive.

Soumis à l'emprise d'un faisceau de contraintes, deux niveaux de décisions conditionnent les flux de trésorerie engendrés par l'exploitation :

- un niveau tactique de décisions, d'une part, avec le choix de la structure du cycle d'exploitation, décisions de court terme à effets à long terme, qui assurent la continuité et la stabilité des flux, par la constitution et la gestion permanente de volants régulateurs entre flux d'exploitation réels et monétaires que sont les stocks, le crédit-client, le crédit-fournisseur, et l'encaisse ;

- un niveau stratégique de décisions, d'autre part, avec le choix de la combinaison productive, décisions de long terme à effets à court terme, qui engagent durablement la vie de l'entreprise, mais qui ont des conséquences quotidiennes sur sa trésorerie à travers le volume, la répartition, le comportement et la couverture des charges.

Dans ce second domaine, plus particulièrement, l'autonomie de décision est bornée, d'abord, par les besoins du marché qui limitent les quantités à produire, ensuite, par la concurrence qui fixe les hauteurs de prix possibles, enfin, par le progrès technologique qui induit le caractère de la combinaison productive compétitive.

Dans la mesure où les encaissements sont générés par l'activité, il semble évident d'affirmer que le risque d'insolvabilité attaché à une combinaison productive dépend du rapport entre coûts fixes et coûts variables. Bien que le renouvellement de « l'analyse des coûts basée sur les activités » mette en relief que les coûts ne dépendent pas du seul facteur « chiffre d'affaires », du point de vue de la gestion de la trésorerie c'est bien la production vendue qui définie à la fois les recettes et les coûts afférents.

Parce que la combinaison productive retenue détermine la structure des charges, elle gouverne la « liquidité » de l'entreprise. Une activité est « liquide » au moment où tous les coûts, quelles que soient leurs échéances, sont couverts par le revenu de cette activité quelles que soient les dates de son encaissement, pour un cycle d'exploitation donnée. Mais tous les flux d'exploitation n'engendre pas de flux de trésorerie, c'est le cas principalement des « charges calculées ». Aussi doit-on distinguer entre « liquidité immédiate » et « liquidité à terme ». La « liquidité immédiate » d'une opération est la trésorerie nette susceptible d'être générée au cours de la période, autrement dit sa solvabilité potentielle. La « liquidité à terme » d'une opération est la trésorerie nette susceptible d'être dégagée au cours de la période, après s'être assuré du maintien de la valeur du capital, autrement dit sa rentabilité. On reconnaît dans la liquidité à terme l'approche du seuil de rentabilité. Du point de vue de la liquidité de la firme, en effet, la capacité d'autofinancement représente la variation potentielle de la trésorerie en fin de période. En d'autres termes, avec une trésorerie nulle en début de période, la capacité d'autofinancement est la « *trésorerie virtuelle* » de fin de période. L'écart entre l'encaisse potentielle et l'encaisse réelle tient aux décalages temporels entre les flux d'exploitation et les flux d'encaisse[1]. La rentabilité, comme on le sait, est le gage de la solvabilité à terme.

Dans ces conditions, la gestion de la trésorerie, considérée par la Théorie Financière comme un problème subséquent, devient la question fondamentale qui sous-tend toute la gestion financière de la firme. Le seuil de rentabilité qui est en réalité un « *seuil de liquidité à terme* », indique que sur la durée de vie des investissements, à un certain niveau de production annuel, la trésorerie garantit le maintien de la valeur du capital et assure, le cas échéant, le renouvellement des investissements. Si l'on ne retient que les flux d'exploitation susceptibles de se transformer en flux de trésorerie, on définit un second seuil de liquidité, volume d'activité pour lequel les recettes couvrent les coûts fixes décaissables, qui peut être qualifié de « *seuil de liquidité immédiate* », et signifie que telle opération garantit pour un certain niveau de production une trésorerie positive, tout en ignorant la date où elle le deviendra[2].

D'une manière générale toute combinaison productive se caractérisera par son degré de liquidité. L'effet de levier de trésorerie est un outil d'analyse du degré de liquidité d'une combinaison productive. Il mesure la sensibilité de la trésorerie, c'est à dire la variation de la « trésorerie virtuelle » qui résulte, toutes choses égales par ailleurs, de la variation des paramètres de l'activité : le volume et la marge unitaire sur coûts variables. Nous présenterons une étude des composantes de l'effet de levier de trésorerie. Une manifestation originale de l'« l'effet de levier de trésorerie » apparaît lorsque la part des coûts fixes varie par rapport au montant total des charges, dans la mesure où la substitution du capital au travail s'accompagne, essentiellement, d'un gonflement des charges d'amortissement, non décaissables par nature, et d'une diminution des coûts variables unitaires. Ce phénomène peut apparaître à l'occasion d'une transformation de la combinaison productive, ou d'une augmentation de la capacité de production. Le risque d'« illiquidité » dépend de l'évolution et de la maîtrise de ces facteurs. Nous en donnerons une approche.

## I - Les composantes de l'effet de levier de trésorerie.

Le taux de variation de la trésorerie virtuelle découle de la conjonction du taux de variation de la production vendue et du taux de variation de la marge unitaire sur coûts variables. L'effet de levier de trésorerie résulte donc de deux phénomènes interdépendants :

- l'élasticité de la trésorerie par rapport à la variation du volume d'activité, d'une part, et,

- l'élasticité de la trésorerie par rapport à la variation de la marge unitaire sur coûts variables, d'autre part.

[1] Juhel J.-C., « Le seuil de solvabilité, instrument d'analyse financière et modèle de prise de décisions » Revue Française de Comptabilité, n° 262, décembre 1994, pp. 60-69.
[2] Juhel J.-C., « Le modèle du seuil de solvabilité, application pratique », 28 pages, revue « Echanges », mai 1995, n°112.

**A - L'élasticité de la trésorerie par rapport au volume d'activité : $E_{I/Q}$.**

L'élasticité de la trésorerie par rapport par rapport au volume d'activité est le rapport entre la variation relative de la trésorerie et la variation relative de la production vendue.

Ainsi, une élasticité de 2 signifie qu'une augmentation de 1 % de la production entraîne une augmentation de 2 % de la trésorerie virtuelle.

Si l'on écrit que :

« T » est la trésorerie

« Q » est le volume vendu

$$E_{I/Q} = \frac{\Delta T / T}{\Delta Q / Q}$$

Le coefficient d'élasticité varie naturellement selon les niveaux de production. Il est positif après le seuil de liquidité et négatif avant ; dans le cas de non-linéarité des flux il redevient négatif au-delà de l'optimum.

Si l'on écrit que la trésorerie, T, est égale à :

$$T = [(p - v) Q] - f$$

où

« p » est le prix de vente unitaire,
« v » est le coût variable unitaire,
« f » est le montant des charges de structure,
et, « Q », les quantités au niveau desquelles l'élasticité est calculée,

L'élasticité s'écrit :

$$E_{I/Q} = \frac{(p - v) Q}{[(p - v) Q] - f} = \frac{mQ}{(mQ) - f}$$

et, si $m = p - v$ désigne la marge unitaire sur coûts variables :

$$\mathbf{E_{I/Q} = \frac{Q}{Q - (f/m)}}$$

Cette relation permet de calculer l'effet de levier de la trésorerie pour une valeur quelconque de la production et pour une marge donnée.

Mais tous les flux d'exploitation, nous l'avons dit, ne se transforment pas par nature en flux de trésorerie. Les charges calculées, essentiellement[3], à savoir les amortissements et les provisions de la période, sont sans influence sur la trésorerie immédiate bien qu'à terme par le biais des exigences de la protection et du renouvellement du patrimoine elles jouent sur le degré de liquidité.

[3] Il en est de même des produits « fixes » non encaissables, tels que « les transferts de charges » lorsqu'il s'agit de transferts à un compte de bilan.

Deux indicateurs sont donc à prendre en considération : *l'élasticité de la trésorerie à terme* ou « effet de levier d'exploitation » - indicateur bien connu - et, *l'élasticité de la trésorerie immédiate* que l'on pourrait qualifier d'« effet de levier d'encaisse ». Dans le premier cas les coûts fixes « **f** » englobent les charges calculées, dans le deuxième cas ils ne comprennent que les charges décaissables.

Effet de levier d'exploitation et effet de levier d'encaisse mesurent la sensibilité de la trésorerie à toute variation du volume des ventes, pour un niveau de production déterminé et pour une marge unitaire donnée. Plus le coefficient de levier est fort plus le degré de sensibilité de la trésorerie est élevé, et attache à toute évolution de l'activité décidée ou subie, un risque important de détérioration ou d'amélioration soit de la liquidité à terme - ou rentabilité, soit de la liquidité immédiate - ou solvabilité potentielle, soit des deux.

L'exemple suivant illustre cette première approche de l'effet de levier de trésorerie. Une entreprise susceptible d'écouler au cours d'une période 2 400 000 unités au prix de 20 € pièce, pour un coût variable unitaire de 12 €, et compte tenu du rapport entre charges de structure décaissables, 2 000 000 €, et charges calculées, 6 000 000 €, estime qu'en automatisant sa production, elle augmentera son effet de levier d'exploitation, sans courir le danger d'une crise de trésorerie. Le graphique 1 présente le seuil de liquidité à terme, ou seuil de rentabilité, et le seuil de liquidité immédiate.

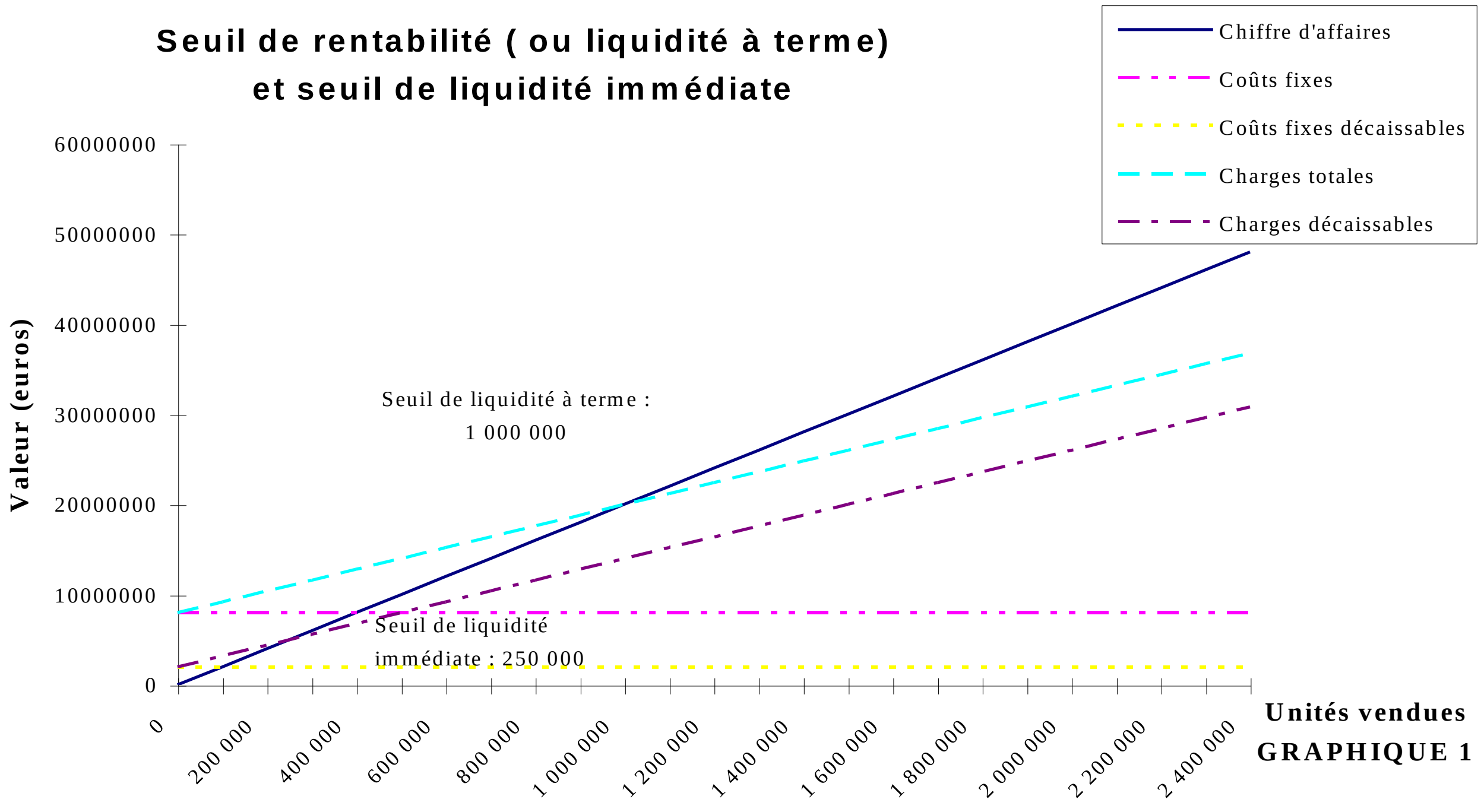

**GRAPHIQUE 1**

Le seuil de liquidité immédiate indique la production critique qui assure une encaisse potentielle positive. Le seuil de liquidité à terme donne la production pour laquelle le maintien de la valeur du capital est garanti. Remarquons que le seuil de liquidité immédiate est par définition toujours antérieur au seuil de rentabilité.

Le graphique 2 représente l'évolution du coefficient de l'effet de levier d'exploitation et celle du coefficient de l'effet de levier d'encaisse pour tous les niveaux de production possibles. Le coefficient de levier d'encaisse est indéfinie pour un niveau de production plus faible que celui de levier d'exploitation. Autrement dit, la liquidité immédiate est plus rapidement moins sensible aux variations de l'activité que la liquidité à terme.

Il convient d'observer, en outre, que l'élasticité de la trésorerie par rapport au volume d'activité n'est déterminante que dans la zone proche des points critiques. Comme on le constate à la lecture du graphique 2, la valeur de l'élasticité diminue très rapidement en dehors de cette zone. Ainsi, au-delà des seuils elle tend rapidement vers 1. En d'autres termes, la variation de la trésorerie devient proportionnelle aux quantités vendues lorsque celles-ci s'éloignent des seuils de liquidité par valeurs positives. La trésorerie à terme, comme nous venons de le remarquer, a une zone de sensibilité plus difficile à quitter que la trésorerie immédiate car se situant à un niveau de production beaucoup plus élevé. En l'occurrence, la liquidité immédiate atteint son seuil pour 250 000 unités vendues, alors que la liquidité à terme n'y parvient qu'à 1 000 000 d'unités. Pour une production prévisionnelle de 2 400 000 unités, l'effet de levier d'encaisse est de 1,12 et l'effet de levier d'exploitation de 1,71.

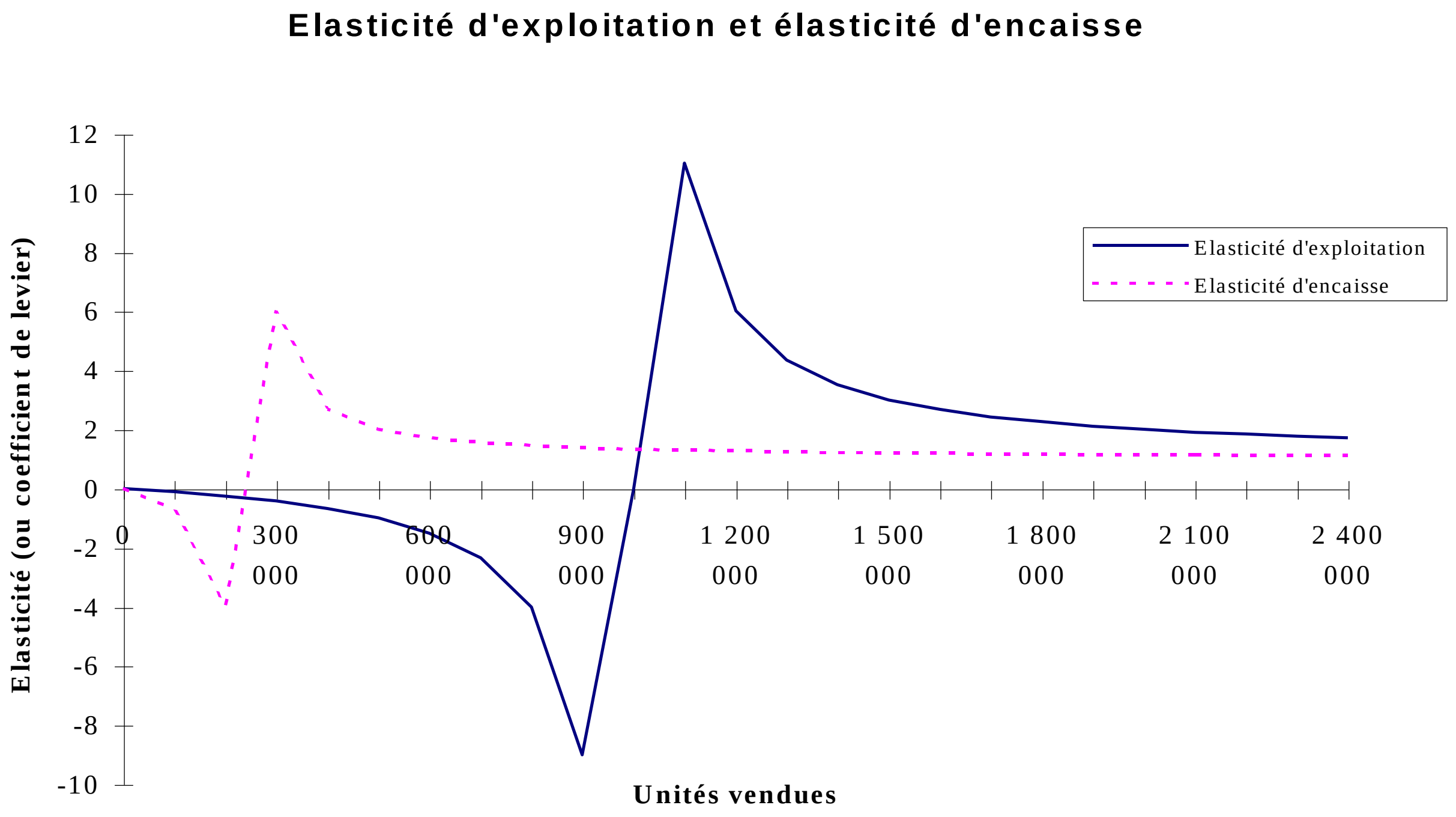

**GRAPHIQUE 2**

Cette remarque trouve sa justification si l'on considère le coefficient de levier de trésorerie précédemment établi :

$$E_{l/Q} = \frac{m\,Q}{(m\,Q) - f}$$

En effet, on peut écrire que lorsque la production, Q, augmente, le rapport f / Q tend vers zéro, et E tend vers 1 :

$$E_{l/Q} = \frac{m\,Q}{Q\,[m - (f/Q)]} \sim \frac{mQ}{mQ} = 1$$

Le tableau suivant d'« Evolution de la *liquidité à terme* et de la *liquidité immédiate* attachées à une structure de coûts » présente les données et les résultats attendus de trois projets alternatifs pris comme exemple. Le projet « 1 » que l'on pourrait qualifier de « sans changement » est la situation initiale. Il reprend les valeurs de l'exemple précédent. L'évolution envisagée de la combinaison productive se caractérise par une augmentation des coûts fixes accompagnée d'une baisse des coûts variables. La sensibilité de la trésorerie se trouve affectée par les changements des paramètres « f » et « m », c'est à dire des coûts variables « v » par rapport aux coûts fixes « f », et de la structure même des charges fixes. Il conviendra d'étudier la nature de ces changements de paramètres et leurs conséquences sur la liquidité de la firme.

***Tableau d'évolution de la « liquidité à terme »***
***et de la « liquidité immédiate » attachées à une structure de coûts***

| PROJETS | 1 | 2 | 3 |
|---|---|---|---|
| Coûts fixes totaux | 8 000 000 | 15 000 000 | 16 800 000 |
| Charges calculées | 6 000 000 | 12 000 000 | 14 400 000 |
| Coûts fixes décaissables | 2 000 000 | 3 000 000 | 2 400 000 |
| Coûts variables unitaires | 12 | 10 | 8 |
| Marge unitaire | 8 | 10 | 12 |
| ***Seuil de liquidité à terme (rentabilité)*** | ***1 000 000*** | ***1 500 000*** | ***1 400 000*** |
| ***Seuil de liquidité immédiate*** | ***250 000*** | ***300 000*** | ***200 000*** |

Le graphique 3 représente les courbes d'évolution des coefficients de levier d'exploitation (qualifié parfois d'opérationnel) et d'encaisse de ces projets. A l'aide de cette information l'entrepreneur peut alors envisager de modifier son activité en connaissant les risques induits probables d'insolvabilité, compte tenu de la profitabilité attendue.

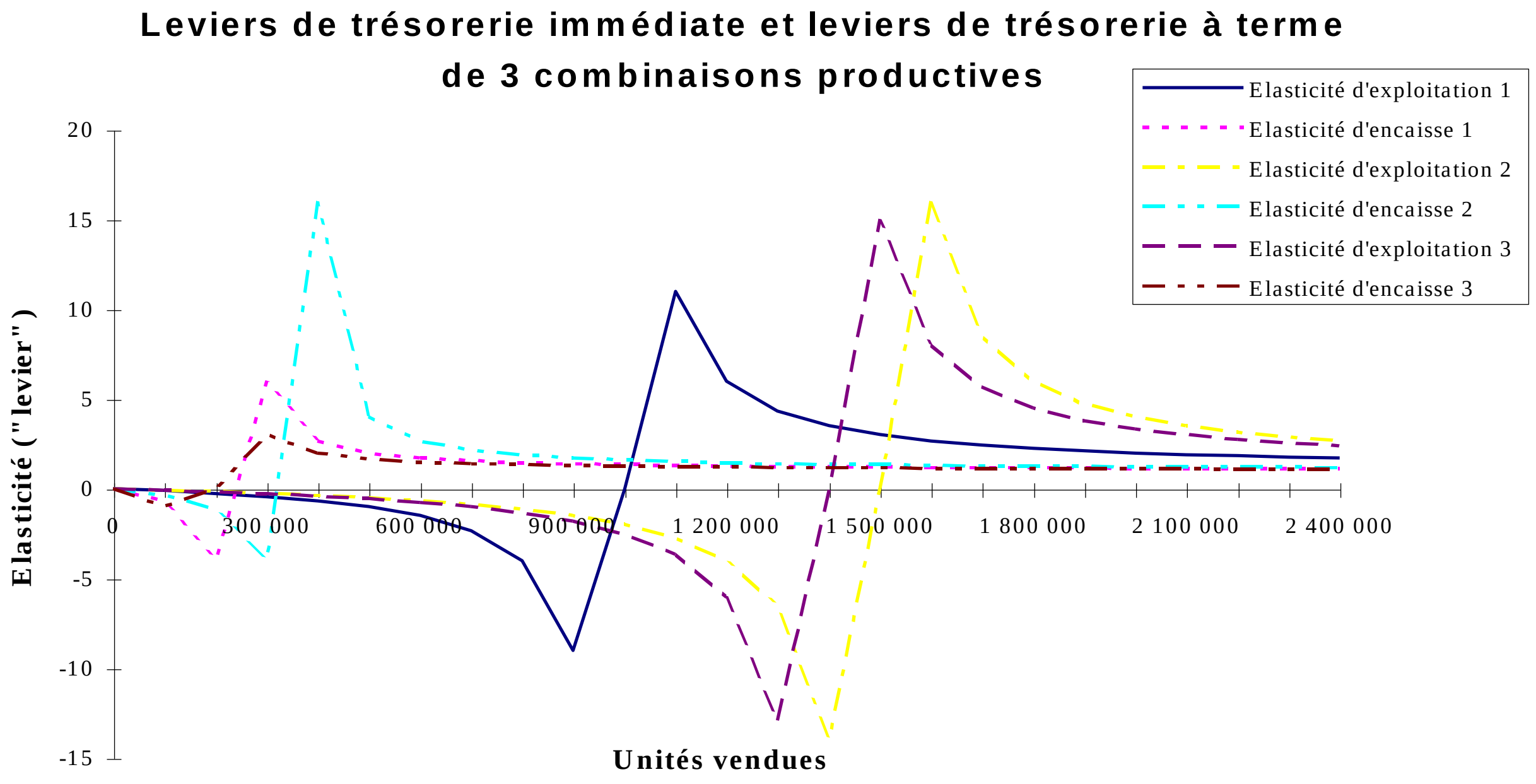

**GRAPHIQUE 3**

On observe que l'intérêt de chaque projet dépend du point de vue où l'on se place. Le tableau « *Performances des projets* » permet de juger de l'opportunité de « moderniser » la combinaison productive, compte tenu d'une capacité de production inchangée. Sur le plan de la sensibilité de la trésorerie immédiate les trois projets sont équivalents. Par contre, en ce qui concerne la liquidité à terme le projet 1 est le plus intéressant car le plus stable. Si le projet 3 génère le bénéfice le plus important en valeur absolue sa rentabilité économique est moitié plus faible que le projet 1. Le projet 2 cumule la plus forte sensibilité de trésorerie à terme et la plus faible profitabilité. Par contre, si la capacité de production augmente avec la modernisation de l'appareil productif les conclusions sont à reconsidérer. Tout accroissement de production atténue les différences de sensibilité de la trésorerie des projets.

## *Performances des projets*

| **Projets** | **1** | **2** | **3** |
|---|---|---|---|
| **Durée de vie de l'investissement** *(année)* | 10 | 10 | 10 |
| **Capacité de production** | 2 400 000 | 2 400 000 | 2 400 000 |
| **Coûts fixes totaux** | 8 000 000 | 15 000 000 | 16 800 000 |
| **Charges calculées** *(amortissement annuel)* | 6 000 000 | 12 000 000 | 14 400 000 |
| **Coûts fixes décaissables** | 2 000 000 | 3 000 000 | 2 400 000 |
| **Capital investi** *(amortissement annuel × durée de vie de l'investissement)* | 60 000 000 | 120 000 000 | 144 000 000 |
| **Marge unitaire** | 8 | 10 | 12 |
| **Marge totale** | 19 200 000 | 24 000 000 | 28 800 000 |
| **Bénéfice** | 11 200 000 | 9 000 000 | 12 000 000 |
| **Rentabilité** *(bénéfices/Capital investi)* | 0,19 | 0,08 | 0,08 |
| **Levier de trésorerie immédiate** | 1,12 | 1,14 | 1,09 |
| **Levier de trésorerie à terme** | 1,71 | 2,67 | 2,4 |

L'effet d'amplification par rapport au volume de l'activité, du levier de trésorerie est donc un phénomène relatif complexe, loin de justifier en toute circonstance un remplacement du travail par le capital qui en gonflant les coûts fixes fragilise la liquidité de l'entreprise, sans apporter une amélioration systématique de la rentabilité. A la lecture du graphique 4, qui reprend les données de l'exemple précédent, on mesure les conséquences de l'effet d'amplification du levier de trésorerie sur la trésorerie immédiate, la Capacité d'Autofinancement, et sur la trésorerie à terme, le Résultat. Le seuil de liquidité à terme d'une opération dépend des coûts fixes totaux, et donc si leur part est importante, des charges calculées, tandis que le seuil de liquidité immédiate ne dépend que des coûts fixes décaissables.

**Trésorerie immédiate et trésorerie à terme en fonction de l'activité et de la structure des coûts**

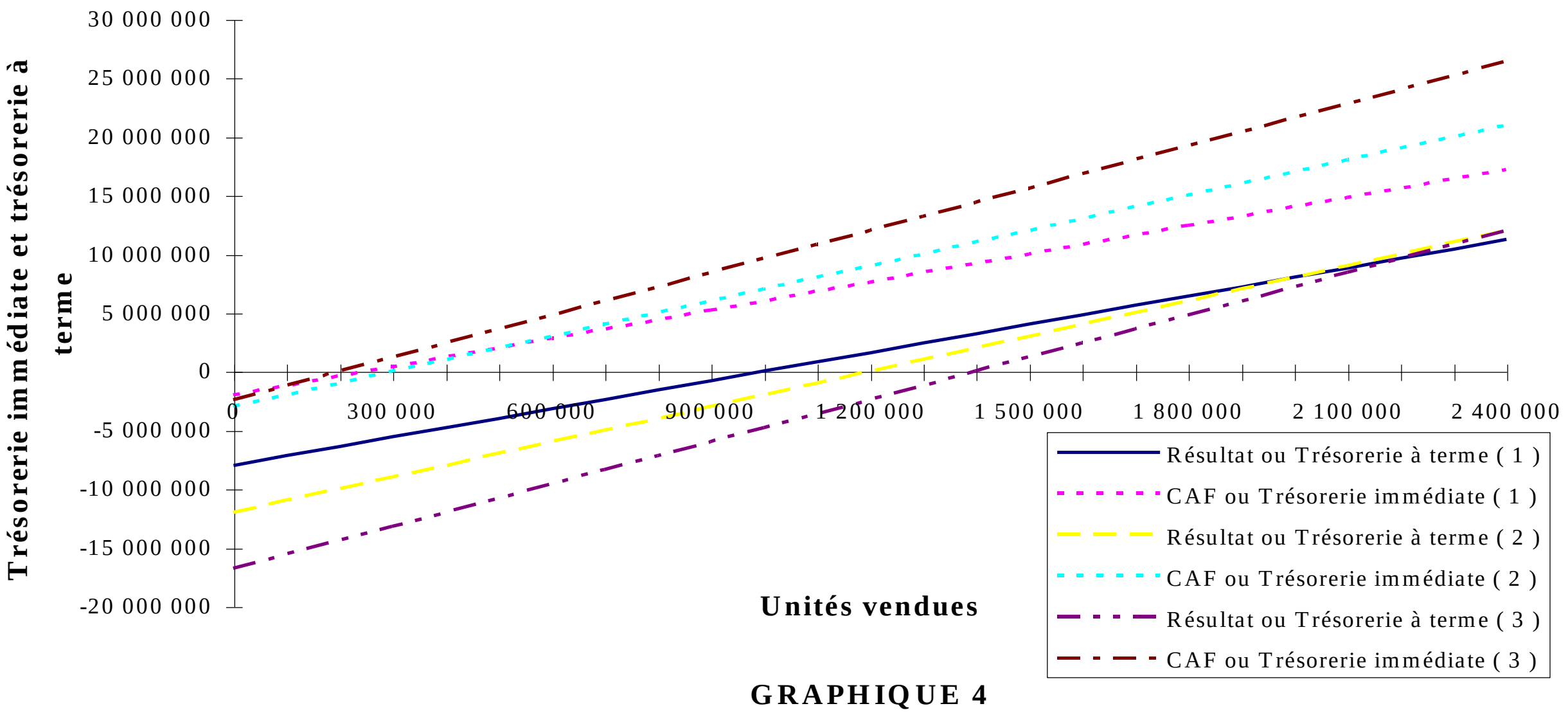

**GRAPHIQUE 4**

Une généralisation de l'approche de l'élasticité de la trésorerie par rapport à la production nous permet de mettre en évidence les limites de l'effet d'amplification. Quelle que soit l'entreprise l'effet d'amplification a la même configuration bien que lié au montant des coûts fixes et au volume d'activité. Au seuil de liquidité immédiate ou à terme, lorsque la marge totale couvre les coûts fixes afférents, l'élasticité tend vers l'infini.

En effet, pour $mQ = f \Rightarrow E_{I/Q} = \dfrac{mQ}{mQ - f} \rightarrow \infty$ , donc

$$\Rightarrow \quad (mQ) - f = 0$$

d'où $\mathbf{Q^* = f / m}$

La production critique est donc bien égale à **f/m**. Le graphique suivant montre le comportement général du coefficient d'élasticité par rapport à la production.

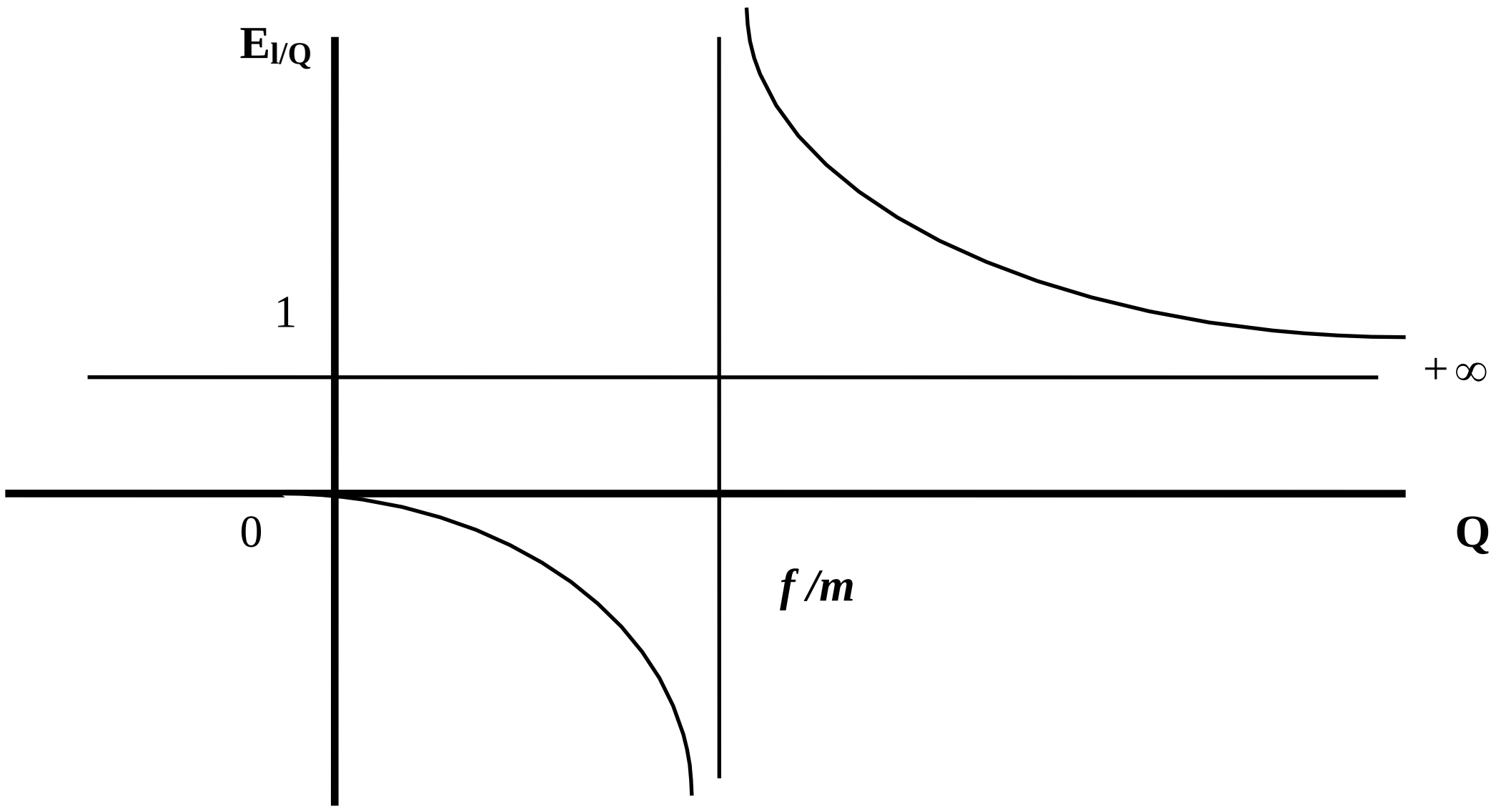

Les valeurs que peut prendre l'élasticité en fonction de la production sont résumées dans le tableau suivant :

| Q | 0 | ½ f/m | 2/3 f/m | f/m | f/m | 2 f/m | 3 f/m | ∞ |
|---|---|---|---|---|---|---|---|---|
| E | 0 | -1 | -2 | - ∞ | + ∞ | 2 | 3/2 | 1 |

On constate qu'au-delà de deux fois le seuil, l'élasticité passe au-dessous de 2 et au-dessous de 1,5 après 3 fois le seuil. Les entreprises qui ont une production de masse auront une trésorerie beaucoup moins sensible que celles produisant à faible échelle, et ce, quelle que soit leur taille. La sensibilité de la trésorerie s'apprécie en fonction du niveau de production que l'on peut atteindre. Le commentaire des valeurs du coefficient d'élasticité se situant en deçà du seuil n'offre pas de difficulté particulière. On remarque également que la valeur du seuil est proportionnelle au montant des charges de structures.

En revanche, pour un niveau de charges de structure donné, et pour un niveau de production donné, la sensibilité de la trésorerie est directement conditionnée par la marge. Cette idée donne un sens complémentaire à la relation de l'élasticité précédemment établie. En systématisant cette observation on peut définir le concept de *« marge critique »*.

### B - L'élasticité de la trésorerie par rapport à la marge unitaire sur coûts variables : $E_{l/m}$.

Pour une production donnée, la marge unitaire sur coûts variables critique est celle pour laquelle il n'y a ni trésorerie virtuelle positive ni trésorerie virtuelle négative. Si la *« marge critique »* est notée m*, elle est égale à f / Q.

En effet,

$$E_{l/m} = \frac{m\,Q}{(m\,Q) - f} = \frac{\mathbf{m}}{\mathbf{m - (f/Q)}}$$

et, E tend vers l'infini si ( mQ - f ) = 0

donc **m* = f / Q**

Le graphique suivant présente, pour un niveau de production donnée, la *marge critique de liquidité à terme*, ou la *marge critique de liquidité immédiate*, f / Q, selon que l'on retienne ou non les charges calculées. En deçà du seuil, m*, la trésorerie virtuelle est négative, au-delà elle est positive.

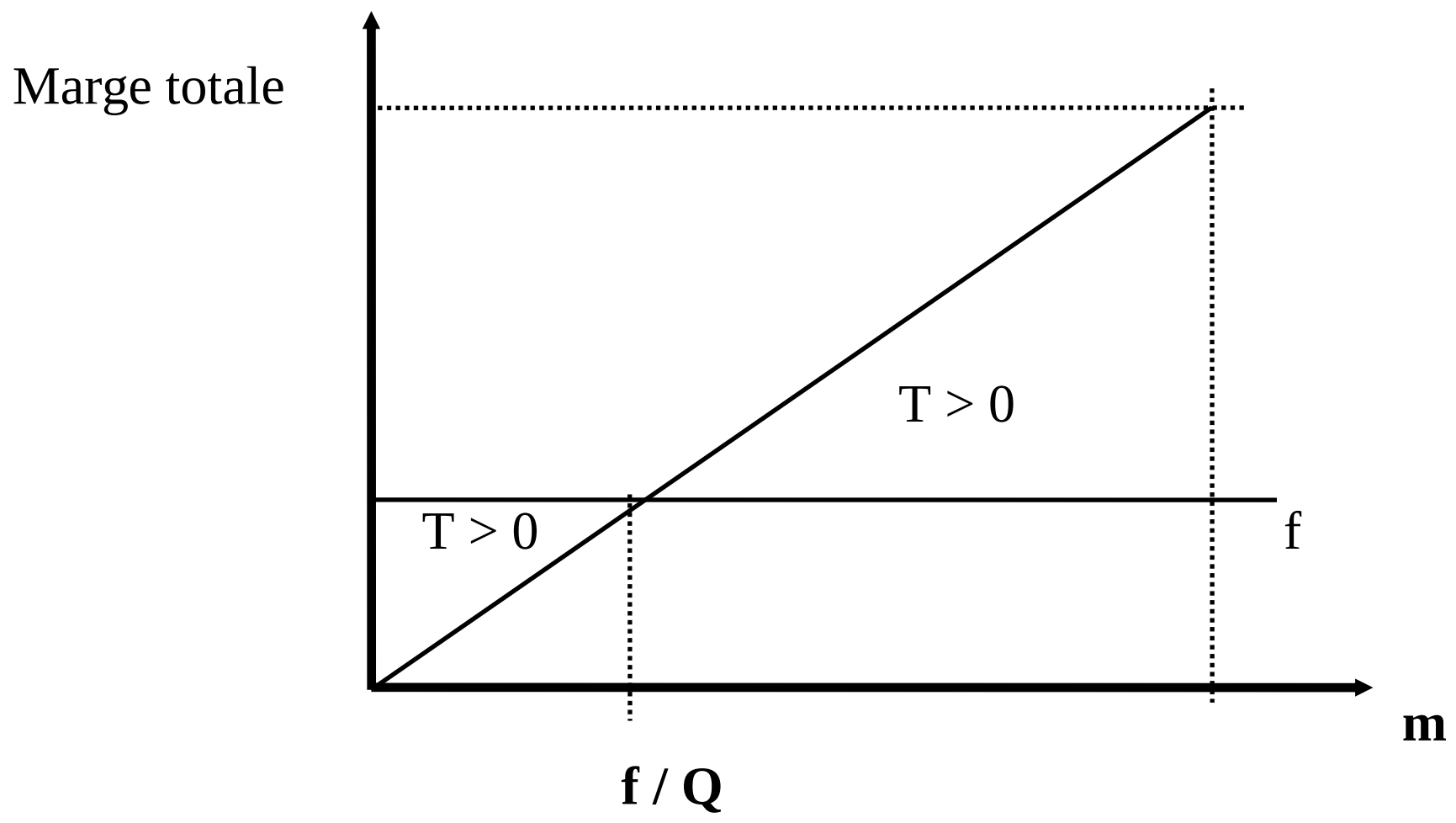

Les coefficients du levier de la liquidité à terme et du levier de la liquidité immédiate par rapport à la marge ont un comportement identique à ceux calculés par rapport à la production, comme le montre le graphique suivant. Quelle que soit l'entreprise l'effet d'amplification a la même configuration bien que lié au montant des coûts fixes et à la marge. Au seuil de liquidité immédiate ou à terme, lorsque la marge totale couvre les coûts fixes afférents, l'élasticité tend vers l'infini.

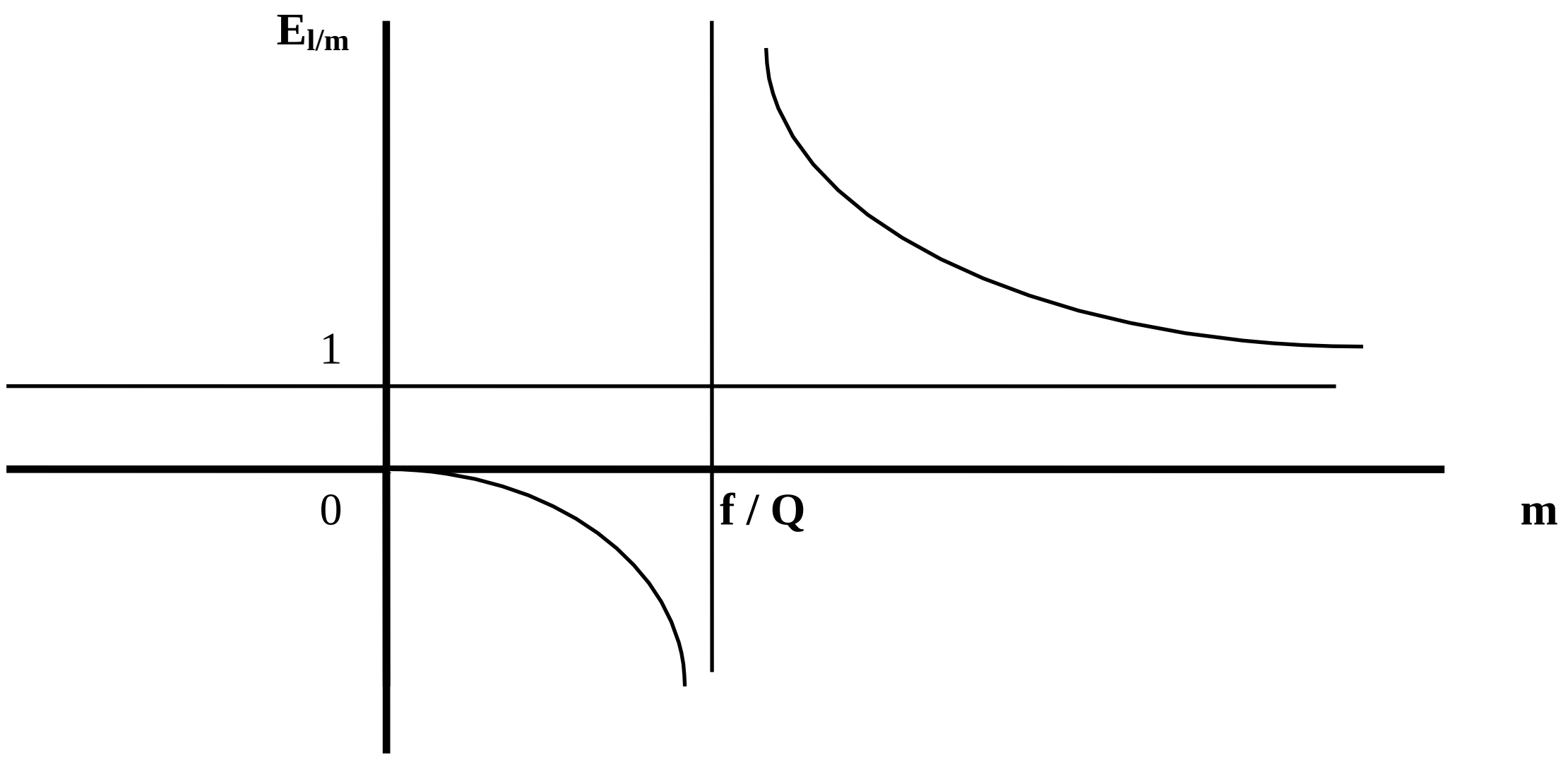

En effet, on peut écrire que lorsque la marge, m, augmente, le rapport f / m tend vers zéro, et E tend vers 1 :

$$E_{l/m} = \frac{m\,Q}{m\,[Q - (f/m)]} \sim \frac{mQ}{mQ} = 1$$

Les valeurs que peut prendre l'élasticité en fonction de la marge sont résumée de la même façon dans le tableau suivant :

| **m** | 0 | ½ f/Q | 2/3 f/Q | f/Q | f/Q | 2 f/Q | 3 f/Q | ∞ |
|---|---|---|---|---|---|---|---|---|
| **E** | 0 | -1 | -2 | - ∞ | + ∞ | 2 | 3/2 | 1 |

Si l'on reprend les données de l'exemple précédent on peut construire le graphique 5 qui donne la valeur des marges critiques que par analogie nous qualifierons de « liquidité à terme » et de « liquidité immédiate » par rapport à la marge.

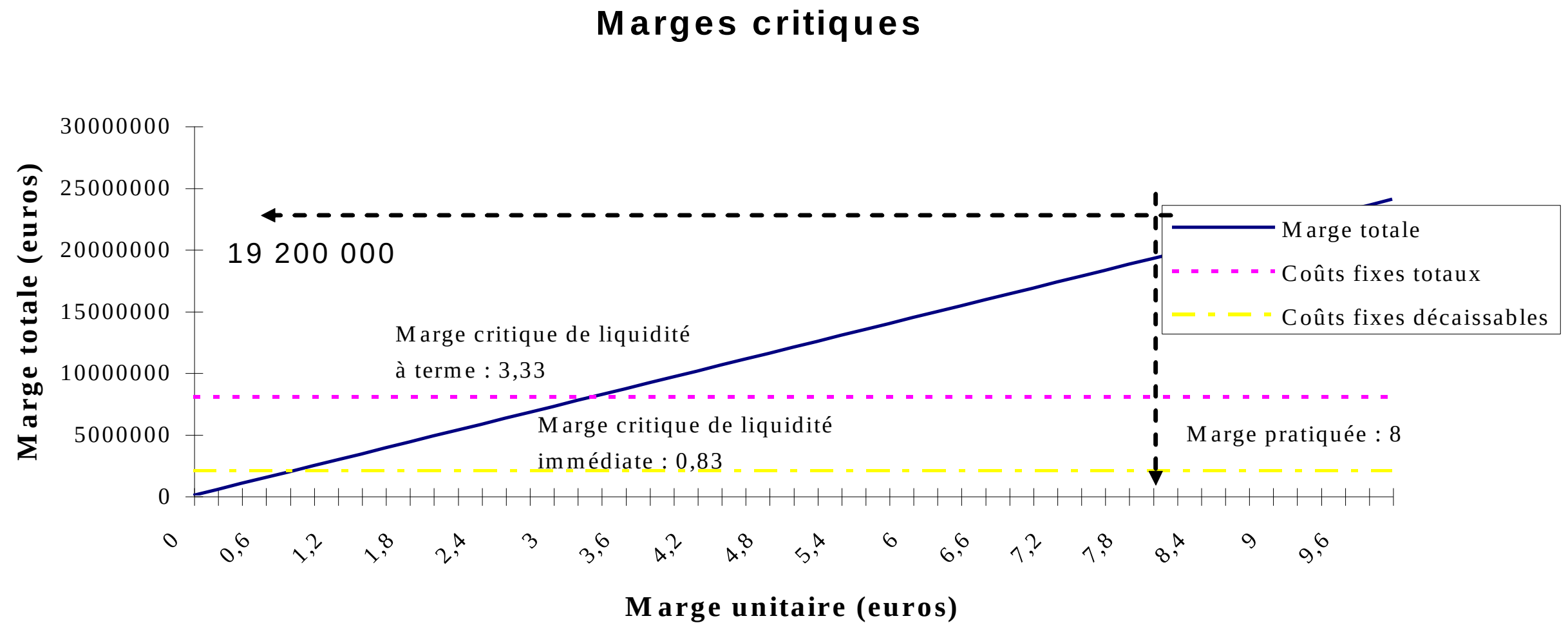

**GRAPHIQUE 5**

Des commentaires similaires à l'analyse présentée précédemment peuvent être faits, notamment en relevant que la marge critique est proportionnelle à la valeur des coûts fixes. De même, l'élasticité de la trésorerie par rapport à la marge n'est déterminante que dans la zone proche des points critiques ; la valeur de l'élasticité diminue très rapidement en dehors de cette zone pour devenir proportionnelle à la marge lorsque celle-ci augmente. La trésorerie à terme a une zone de sensibilité par rapport à la marge critique plus difficile à quitter que la trésorerie immédiate car se situant à un niveau de marge beaucoup plus élevé, comme le montre le graphique 6.

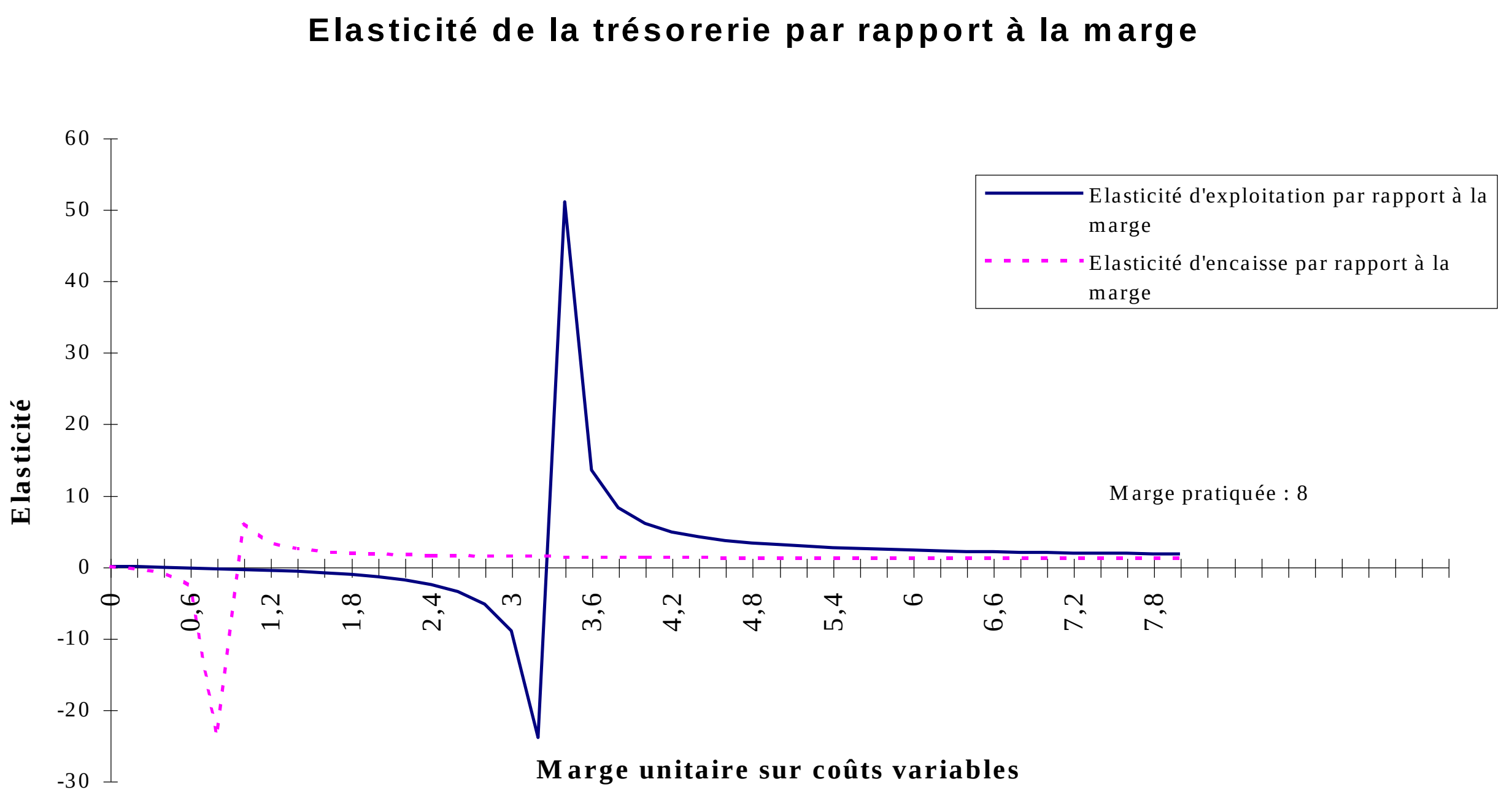

**GRAPHIQUE 6**

En résumé, **les facteurs « f / m » = Q* et « f / Q » = m* conditionnent la sensibilité de la trésorerie de la firme** tant sur le plan de sa liquidité à terme que sur le plan de sa liquidité immédiate. La trésorerie virtuelle est pour ces valeurs la plus sensible à toute modification de la conjoncture managériale : les quantités vendues, les prix du marché et les coûts de production.

Toute stratégie de la trésorerie doit intégrer non seulement le volume de la production mais aussi et pour une même part la marge unitaire sur coûts variables dégagée, et, apprécier le risque d'illiquidité à deux niveaux de charges de structure, coûts fixes totaux et coûts fixes décaissables.

Quelle que soit la nature de la combinaison productive, la sensibilité de la trésorerie qu'une entreprise subie ne joue sérieusement que pour des valeurs relativement faibles de Q et de m par rapport à f. Lorsque ces paramètres de l'exploitation augmentent, pour une valeur inchangée des charges de structure la sensibilité de la trésorerie s'atténue, d'une part, parce que Q et m s'éloignent par valeurs positives respectivement des seuils « f / m » et « f / Q » - beaucoup plus rapidement d'ailleurs, comme nous l'avons vu, pour la liquidité immédiate - et, d'autre part, parce que ces deux rapports tendent vers zéro lorsque m ou Q augmentent, comme le précisent les deux schémas suivants :

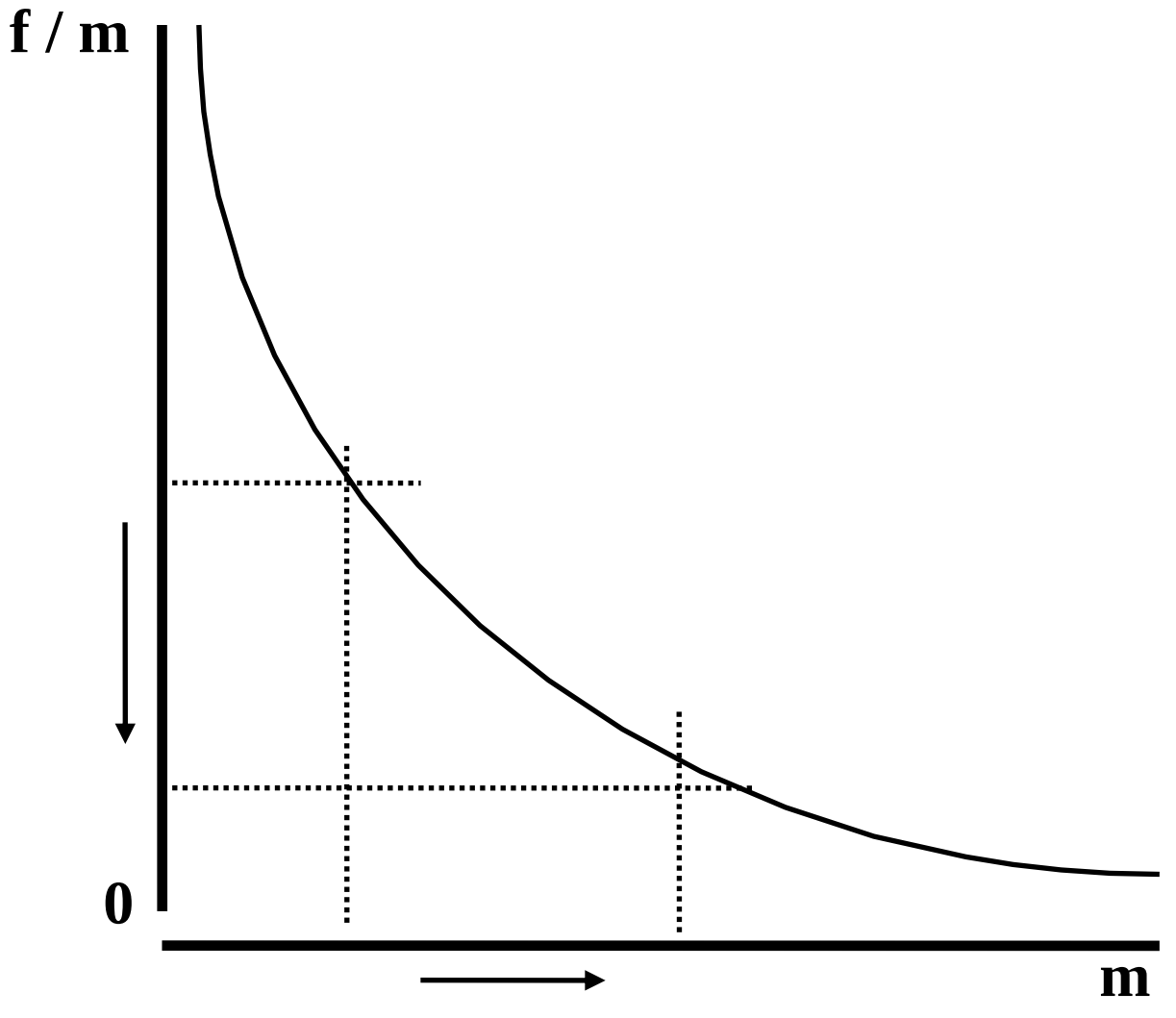

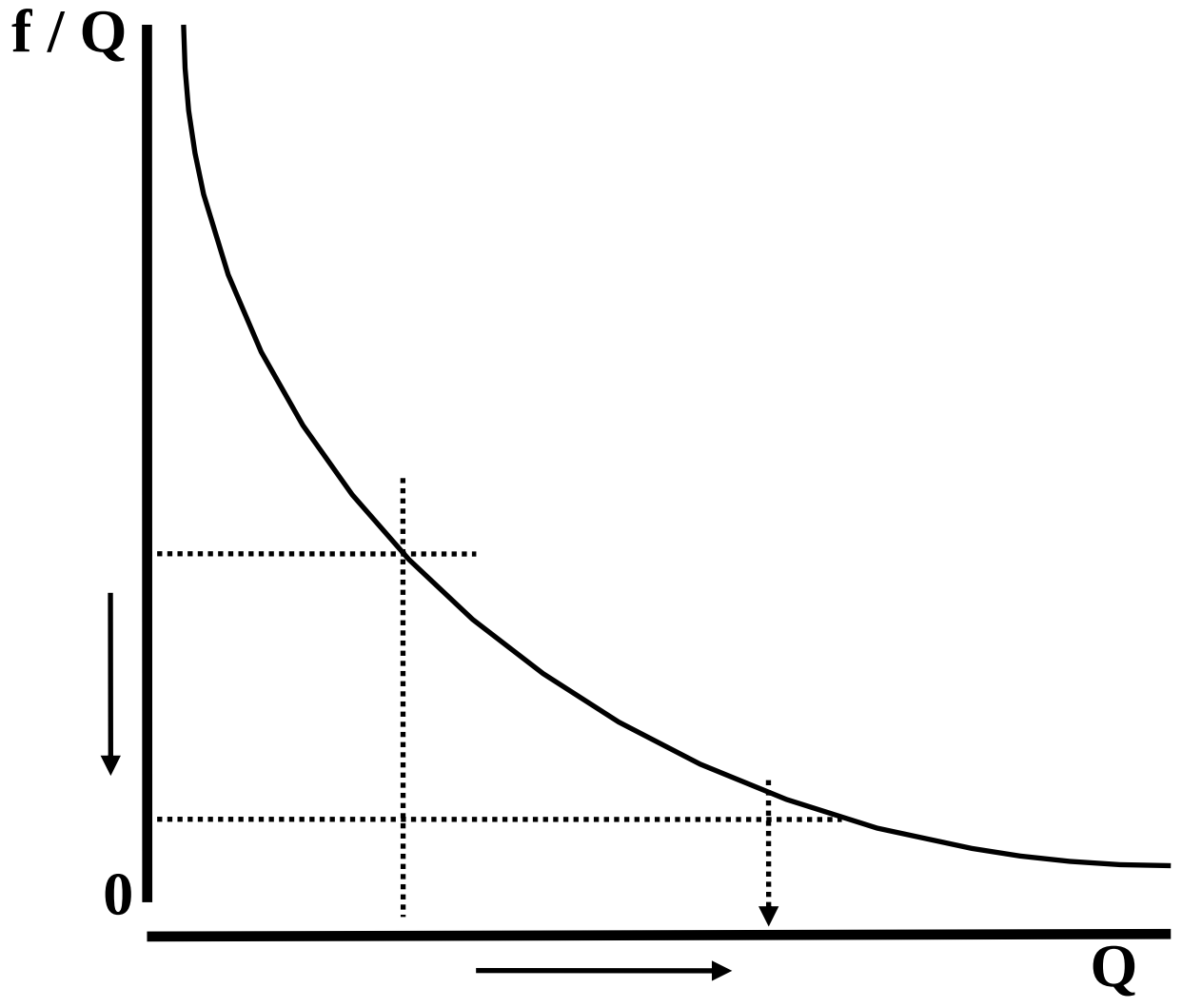

Si l'effet de levier de trésorerie est déterminé par la nature de la combinaison productive, il ne devient facteur stratégique que pour les entreprises à faible capacité de production ou à marge réduite. Le graphique 7 suivant représente l'évolution des seuils de liquidité (immédiate ou à terme, selon le niveau de charges fixes retenues) en fonction de la conjonction du volume de production et de la marge unitaire sur coûts variables confrontée aux charges de structure caractéristiques de la combinaison productive concernée.

Cette illustration reprend les données déjà énoncées de l'exemple précédent.

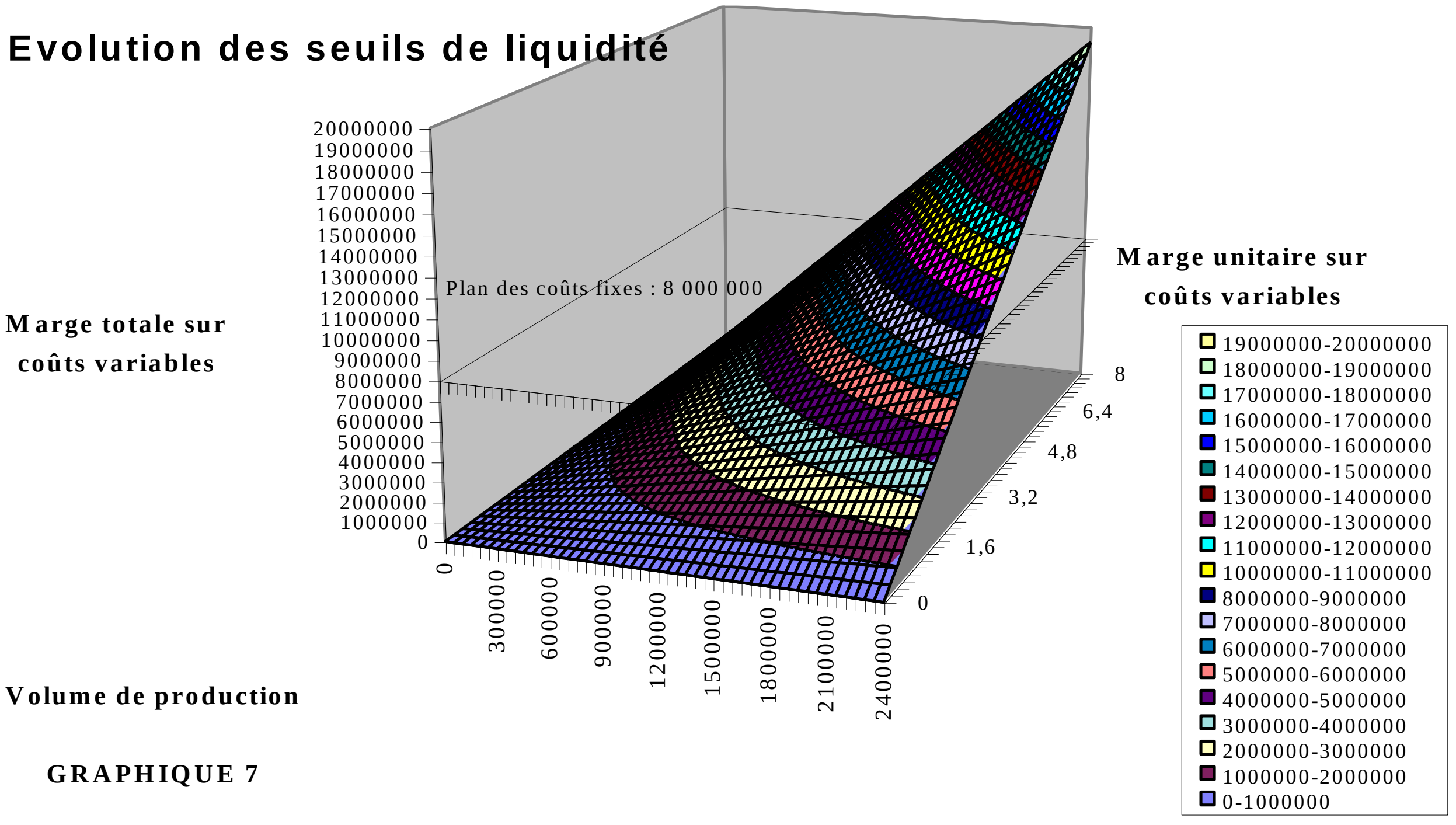

**Niveaux de coûts fixes**

Ainsi, les coûts fixes totaux s'élevant à 8000, on peut lire sur le graphique 7 que pour une marge de 8 le seuil de liquidité à terme est de 1 000 000 unités vendues (limite entre zone verte et violette), et pour une production de 2 400 000 unités la marge critique de liquidité à terme est de 3,33 (limite entre zone verte et violette). Il existe donc des courbes d'indifférence de liquidité déterminées par le niveau des coûts fixes, pour lesquelles la trésorerie virtuelle est nulle, associant un panel de combinaisons productives définies par leur marge ou leur capacité de production et équivalentes du point de vue de la liquidité. Le graphique 8 donne la projection sur un plan et à une échelle plus grande des courbes d'indifférences de liquidité correspondant aux paramètres retenus, et permet de considérer toutes les situations possibles.

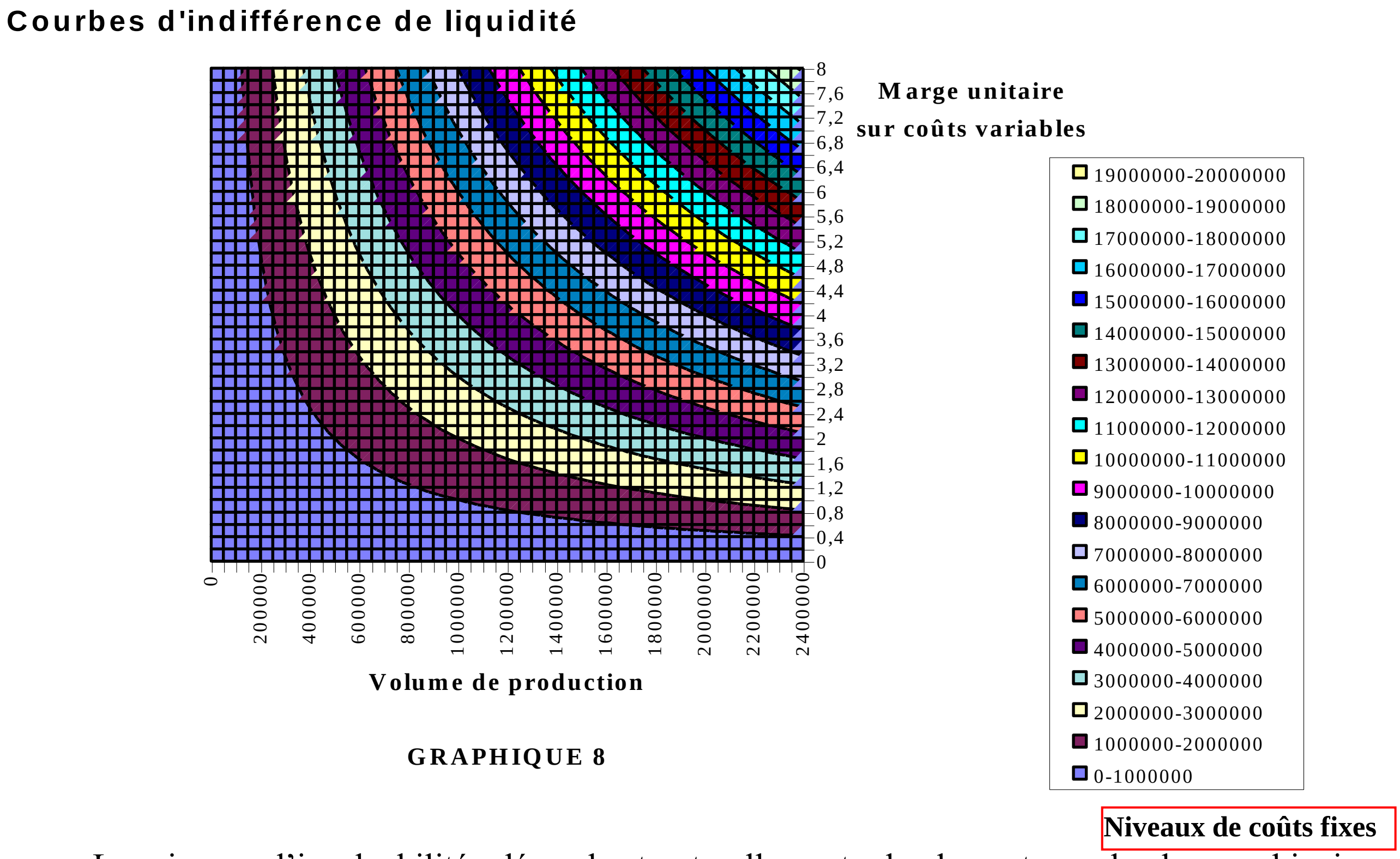

GRAPHIQUE 8

Le risque d'insolvabilité dépend structurellement de la nature de la combinaison productive. L'analyse des conditions de transformation de la combinaison productive permettra de définir le contenu du risque d'insolvabilité.

Nous avons vu que l'effet de levier de trésorerie permet de mesurer la sensibilité de la trésorerie qui se trouve affectée par l'évolution des paramètres « f », « m », et « Q ». Autrement dit la trésorerie virtuelle, immédiate ou à terme, est conditionnée par la réaction des coûts opérationnels « v » à la variation des coûts fixes « f ». En étudiant la nature du comportement de ces paramètres et leurs conséquences sur la liquidité de la firme, on pourra proposer une méthode d'évaluation du risque d'insolvabilité.

## II - Combinaison productive et liquidité de la firme.

Les amplifications à variables multiples du levier de trésorerie jouent dans le cadre de la combinaison productive choisie. Cette dernière se caractérise par sa structure de charges. Toute transformation de la combinaison productive modifie la structure des charges et probablement le risque d'insolvabilité.

Après avoir analysé les paramètres de l'évaluation du risque d'insolvabilité attaché à la nature de la combinaison productive, nous proposerons une méthode d'évaluation du risque d'insolvabilité reposant sur une double approche. En effet, la combinaison productive peut varier indépendamment de la capacité de production par modernisation de l'appareil productif, ou suite à son augmentation résultat d'une politique de croissance interne.

**A - Les données de l'évaluation du risque d'insolvabilité attaché à la nature de la combinaison productive.**

Toute combinaison productive, qu'elle soit « labour-saving » ou « capital-saving », obère le risque d'insolvabilité, mais c'est avant tout son évolution qui est porteuse de risque. Le changement du rapport entre coûts d'activité et coûts de structure transforme l'effet de levier de trésorerie en modifiant les rapports f/m et f/Q. L'élasticité des coûts d'activités par rapport aux coûts de structure est la variable cruciale qui conditionne l'évaluation du risque d'insolvabilité inhérent à tout processus productif.

L'introduction du progrès technologique dans l'exploitation d'une entreprise n'a de sens sur le plan financier que si toute augmentation des coûts fixes s'accompagne d'une diminution des coûts variables, ou plus précisément d'un accroissement de la marge unitaire sur charges d'activité. Si l'on veut exprimer que la variation relative des charges fixes globales s'accompagne d'une variation relative inverse des charges d'activité, on peut noter :

« v » est le montant des coûts variables unitaires, et « $\Delta v$ » leur variation,
« f » est le montant des coûts fixes totaux, et « $\Delta f$ » leur variation,

et, *l'élasticité des coûts variables par rapport aux coûts fixes* s'écrit :

$$E_{v/f} = \frac{\Delta v / v}{\Delta f / f} < 0$$

Le rapport est négatif puisqu'un accroissement des charges de structure engendre d'une baisse relative des charges unitaires d'activité.

Ce phénomène est confirmé par l'observation historique du comportement des coûts dans les entreprises. L'évolution des coûts fixes se caractérise par leur importance croissante par rapport au total des charges. Ce fait s'explique d'abord, par la diffusion du progrès technique générateur de gains de productivité, ensuite, parce que le coût du capital est inférieur au coût du travail, et que le premier facteur est moins contraignant à gérer que le second. La mécanisation, l'automatisation et l'informatisation d'un nombre croissant d'opérations ont rendu possibles les économies de coûts sur le « travail ».

Ce faisant les entreprises notamment celles à production limitée ou à marge étroite, fragilisent leur liquidité dans les conditions développées précédemment. Aussi, les aléas de la conjoncture dans de nombreux secteurs professionnels provoquent depuis quelques années un ralentissement, ou plus exactement une transformation de l'évolution. Pour un même volume d'activité les entreprises tentent de baisser leurs coûts de structure pour être moins vulnérables à la récession. Grâce à une nouvelle stratégie de production, en recourant à la sous-traitance parfois « délocalisée », et au personnel intérimaire elles atténuent la sensibilité de leur trésorerie aux revers d'activité. La recherche de procédures technologiques plus performantes reste cependant essentielle afin d'améliorer les marges de productivité, d'augmenter les capacités de production, ou encore de proposer des produits nouveaux pour gagner des parts de marché sur la concurrence.

Par conséquent, le lien fonctionnel dû au comportement de l'entrepreneur, entre les variations relatives des charges reste un principe structurel des organisations productives. Le graphique suivant représente de façon traditionnelle, les valeurs limites que peut prendre le coefficient d'élasticité des coûts variables par rapport aux coûts fixes :

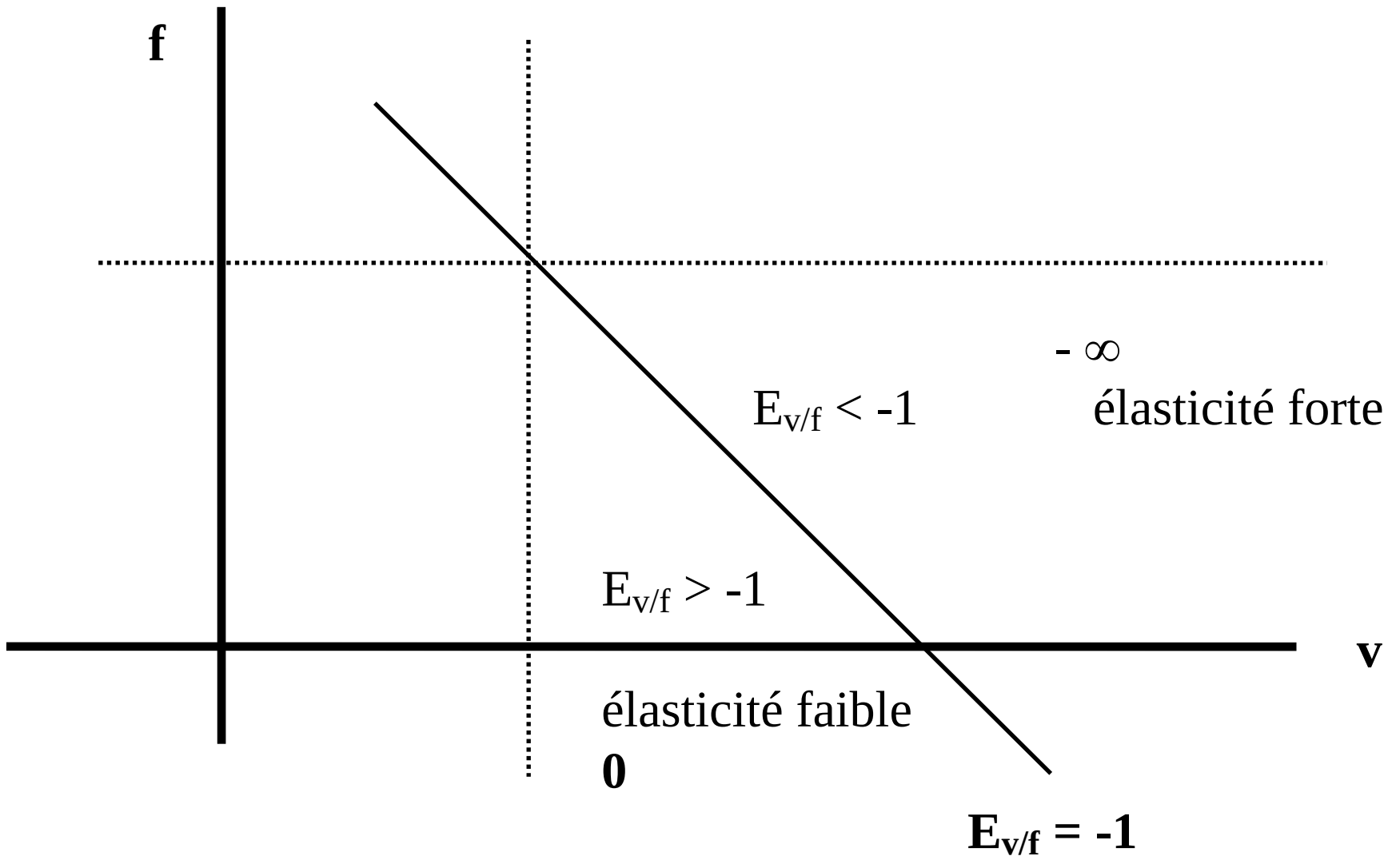

Ce qui signifie que :

si $\Delta v / v > \Delta f / f \Rightarrow E_{v/f} < -1$ l'élasticité est forte,

et, si $\Delta v / v < \Delta f / f \Rightarrow E_{v/f} > -1$ l'élasticité est faible.

Il convient de noter que « v » ne pouvant prendre une valeur négative et « Δv » ne pouvant indiquer qu'une baisse en réaction à l'augmentation de « f », le rapport « Δv /v » ne peut varier que de 0 à - 1, linéairement (si v = Cte) ou non (si Δv = Cte), comme le montre le graphique suivant :

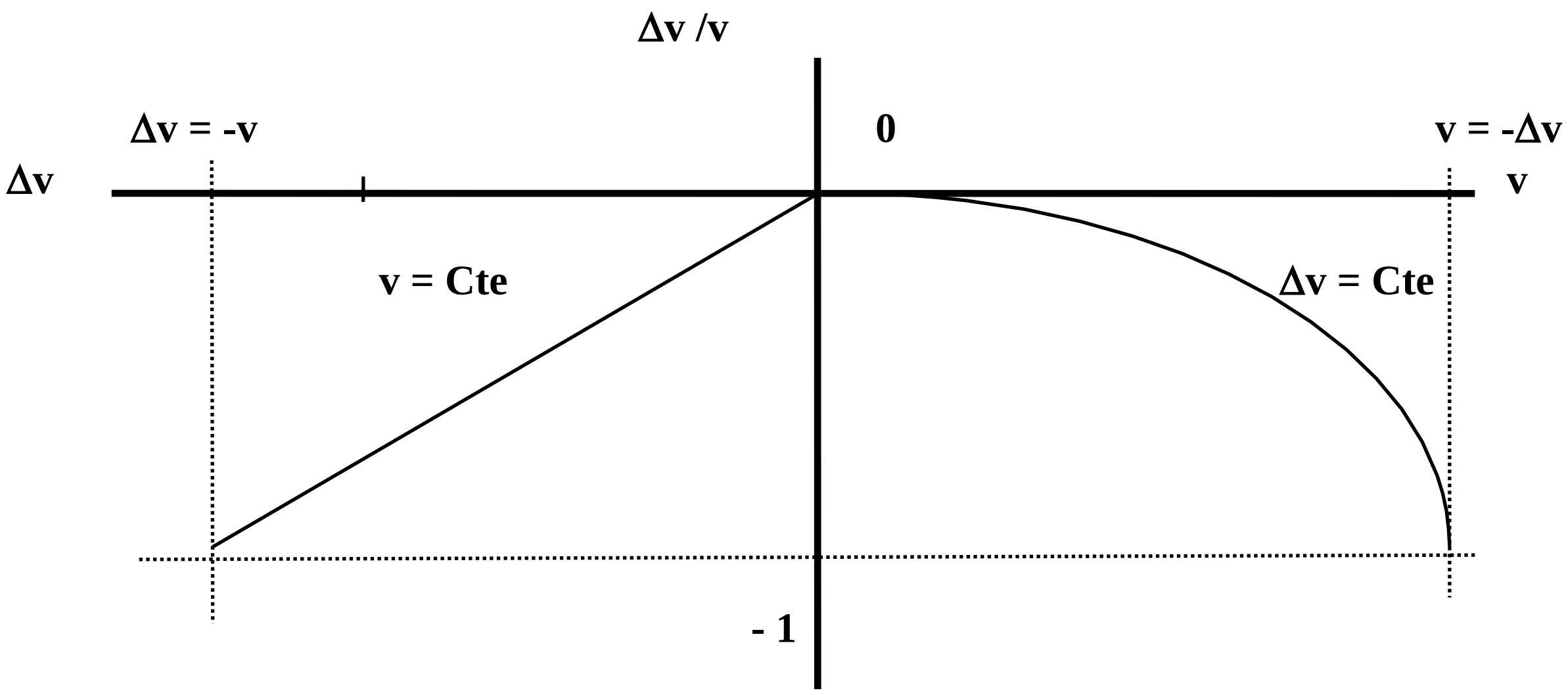

Par contre le taux de variation des charges de structure évolue dans la plage suivante :

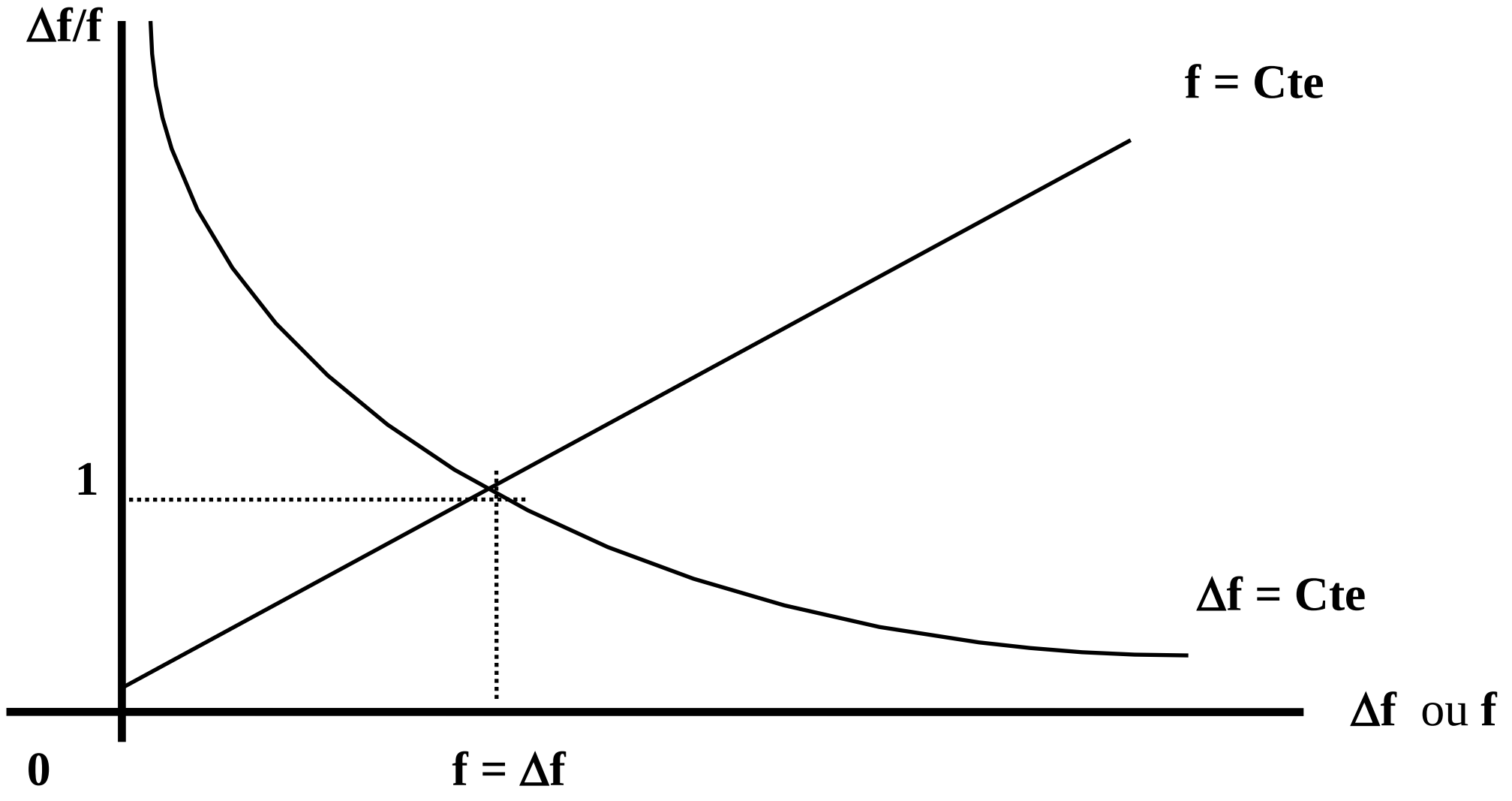

Soit, si f est une constante :

| **Δf** | **0** | ↑ | **f** | ↑ | + ∞ |
|---|---|---|---|---|---|
| **Δf /f** | 0 | ↑ | 1 | ↑ | + ∞ |

et, si Δf est une constante :

| **f** | **0** | ↑ | **Δf** | ↑ | + ∞ |
|---|---|---|---|---|---|
| **Δf /f** | + ∞ | ↓ | 1 | ↓ | 0 |

Donc le taux de variation des coûts fixes « Δf /f » évolue entre 0 et + ∞, selon le comportement de chaque terme.

Ces résultats peuvent être détaillé dans le tableau suivant :

## Plages d'élasticité des coûts variables par rapport aux coûts fixes

| **Δv /v** / **Δf /f** | **0** | ↓ | ↓ | ↓ | ↓ | **- 1** |
|---|---|---|---|---|---|---|
| **0** | ind. | - ∞ | - ∞ | - ∞ | - ∞ | - ∞ |
| ↑ | 0 | - 1 | | *Valeurs inférieures à - 1* | | |
| ↑ | 0 | | - 1 | | | |
| ↑ | 0 | | | - 1 | | |
| | 0 | | | | - 1 | |
| 1 | 0 | | | | | - 1 |
| | 0 | | | | | |
| ↑ | 0 | | | *Valeurs supérieures à - 1* | | |
| ↑ | 0 | | | | | |
| + ∞ | 0 | 0 | 0 | 0 | 0 | 0 |

L'ampleur de la baisse des charges d'activité en réponse à la hausse des charges de structure va modifier de façon spécifique l'effet de levier de trésorerie. A chaque changement de combinaison productive il faut s'attendre à une évolution de la sensibilité de la trésorerie, c'est à dire de $E_{I/Q}$ et de $E_{l/m}$. L'étude du degré et des plages d'élasticité des coûts variables par rapport aux coûts fixes totaux d'une part, et par rapport aux coûts fixes décaissables d'autre part, pour des types de combinaisons productives donnés semble d'un grand intérêt pour apprécier les risques d'insolvabilité liés à une structure d'exploitation.

Si l'on note la relation entre coûts variables unitaires et coûts fixes : **v = af + b**, ( on a retenu par simplification un modèle linéaire, mais un modèle non linéaire ne changerait pas les observations ) où « a » est un coefficient représentatif de la variation de « v » par rapport à « f », qui donc sera négatif par définition, et généralement compris entre - 1 et 0 (sauf cas très particulier de production à très faible échelle), et, « b » la valeur maximale que peut prendre « v » lorsque « f » est à un niveau minimum (nous considérerons pour la facilité de l'étude, lorsque f = 0). De plus, par construction, « af » est toujours inférieur à « b » - en valeurs absolues. Nous pouvons tracer la droite représentative de cette fonction :

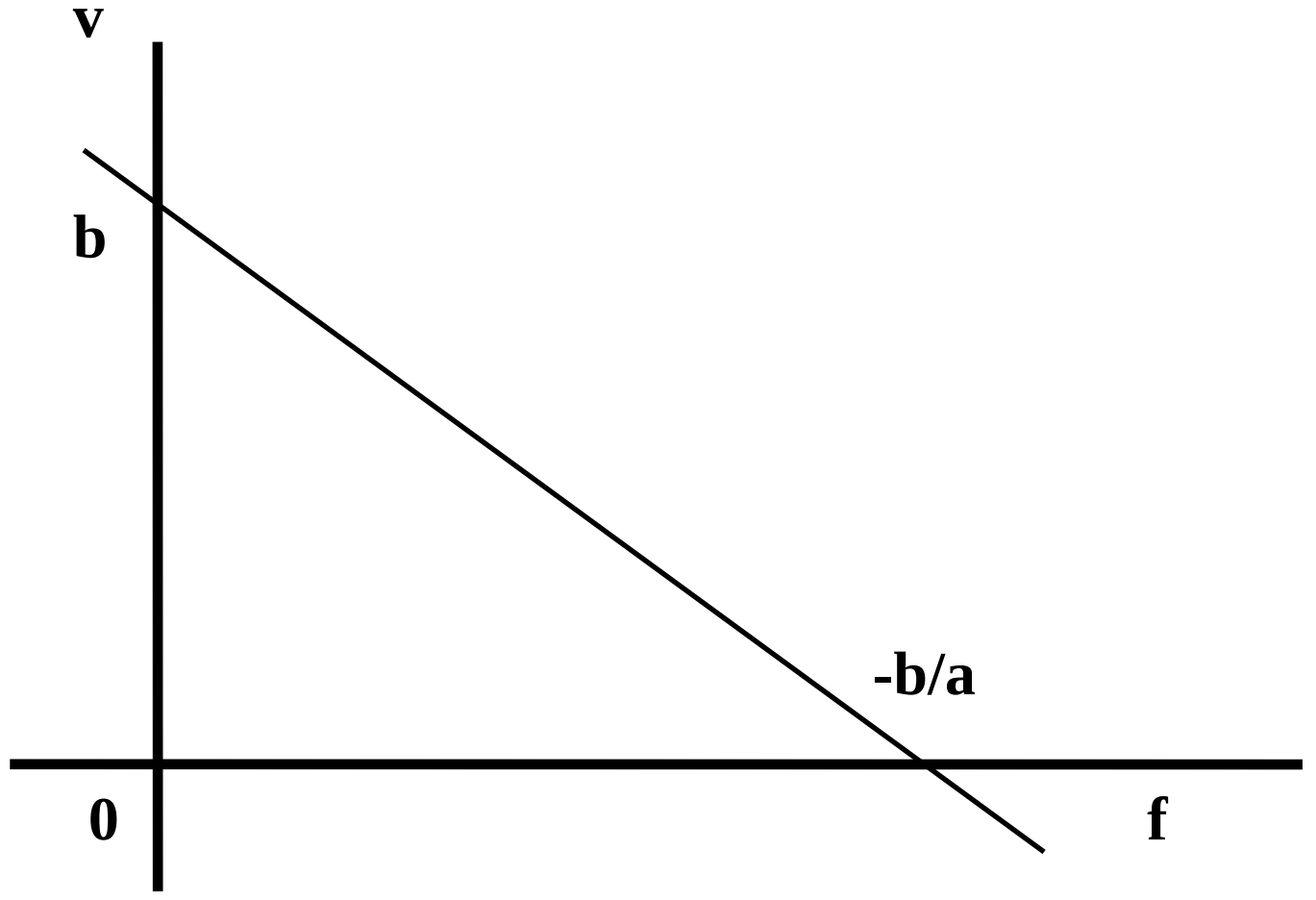

*Pour v = 0 ⇒* ***f = -b/a > 0***
*( puisque* ***a < 0*** *)*

L'élasticité des coûts variables par rapport aux coûts fixes peut donc s'écrire :

$$E_{v/f} = a \frac{f}{af + b}$$

*(où « a » est négatif par construction)*

dont l'évolution se représente par la courbe suivante :

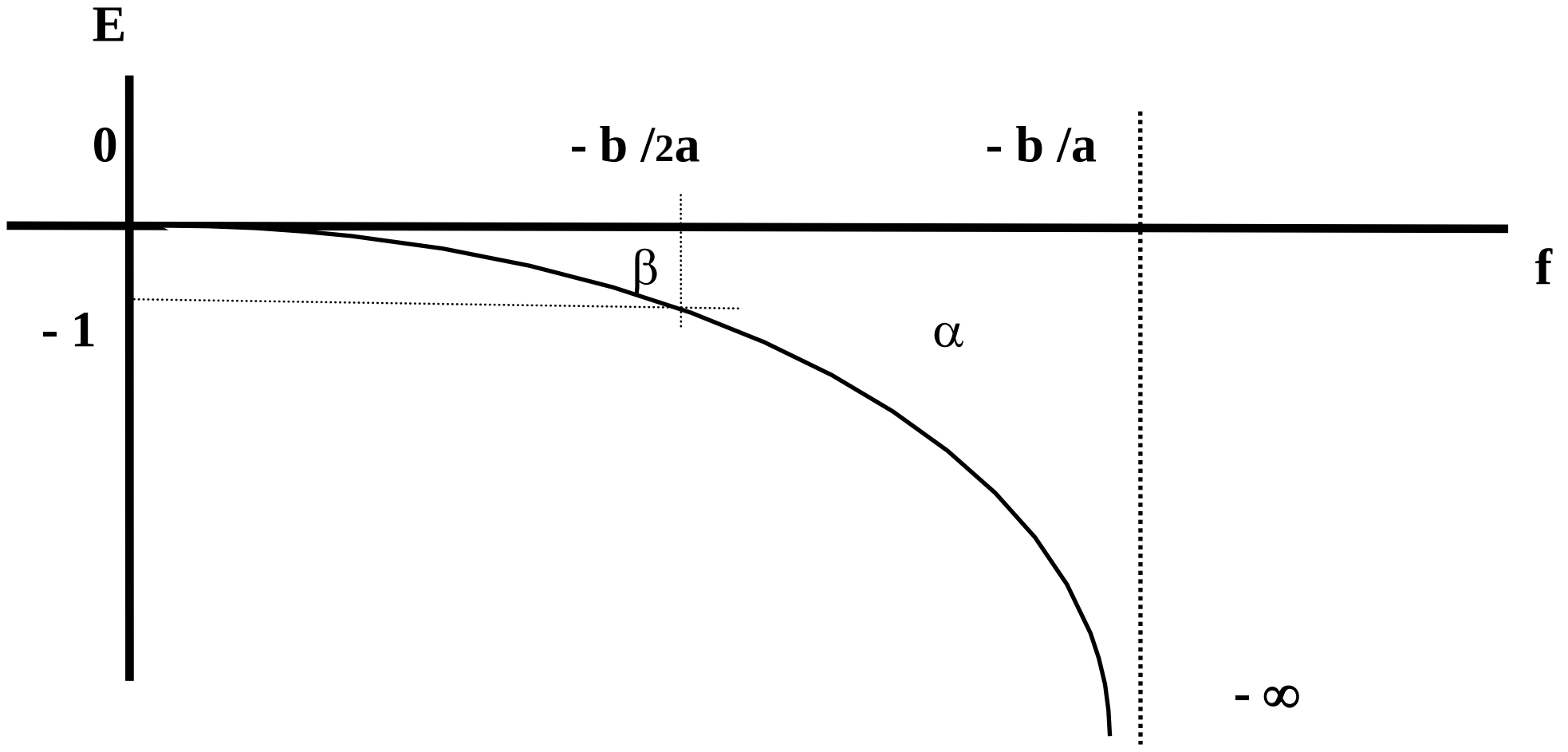

L'interprétation de cette courbe conduit à remarquer que pour toute valeur donnée de coûts fixes existe une limite à l'intervalle de variation de l'élasticité entre zéro et - ∞ pour des valeurs de « f » se situant entre zéro et -b /a. Si l'on néglige les valeurs extrêmes on retient que les variations de « v » par rapport à « f » peuvent se situer aussi bien au-dessus de - 1 (zone α), qu'au-dessous (zone β), selon les contraintes imposées par les conditions technologiques, financières et sociales de l'entreprise, et, le savoir faire de l'entrepreneur. Pour des charges de structure s'élevant à -b/2a l'élasticité sera égale à -1. Au-delà de f = -b/a le problème est économiquement indéterminé ; ce qui ne veut pas dire que les coûts fixes sont limités, mais simplement que les paramètres retenus de l'équation v = af + b ne traduisent pas ou plus le comportement des coûts résultant de la politique de l'entrepreneur. Pour que le rapport de l'élasticité soit négatif il faut, en effet, que le dénominateur soit positif, c'est à dire que af < b - en valeurs absolues.

L'exemple qui suit illustre ces remarques et montre que notre approche rend compte de façon satisfaisante de la réalité. Les données numériques restent celles utilisées précédemment. Nous avons en outre supposé que les coûts fixes pouvaient varier de 1 000 000 à 15 000 000 d'euros et dans le même temps, les coûts variables unitaires de 20 à 6 euros. Les graphiques 9 et 10 donnent une idée de ce que peut être cette relation de l'élasticité des charges d'activité par rapport aux charges de structure, dans le contexte de stratégie de développement déjà défini. Comme nous l'avons vu sur le plan théorique, lorsque la variation relative des coûts variables devient supérieure à la variation relative des coûts fixes - en valeurs absolues - l'élasticité passe au-dessous de -1.

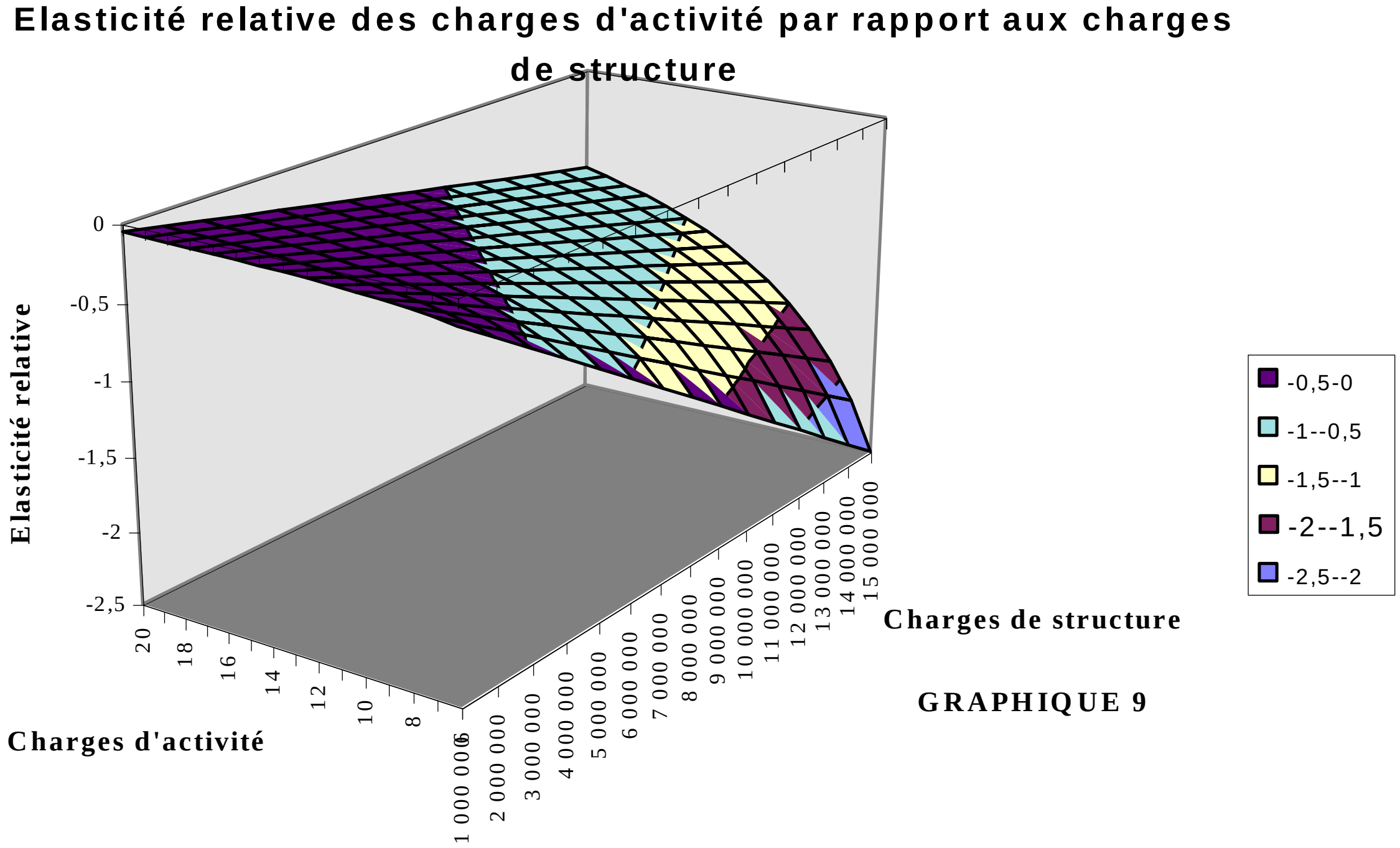

GRAPHIQUE 9

En ce qui concerne la gestion de la trésorerie, on aura compris que pour toute combinaison productive la baisse des charges d'activités par rapport à l'augmentation des charges de structure doive se faire selon un certain chemin du modèle présenté - voir le tableau des « Plages d'élasticité des coûts variables par rapport aux coûts fixes », page 16 - si l'on veut éviter toute modification de la sensibilité de la liquidité. La variation doit être inversement proportionnelle. Une réduction relative plus faible des charges variables se traduirait par une fragilisation de la situation de trésorerie. En effet, toute économie au niveau des coûts d'activité peut être annihilée par un accroissement relatif trop important des charges de structure, ou inversement, l'augmentation des charges de structure peut ne pas s'accompagner d'une diminution suffisante des charges d'activité.

**Elasticité relative des charges d'activité par rapport aux charges de struture ( projection )**

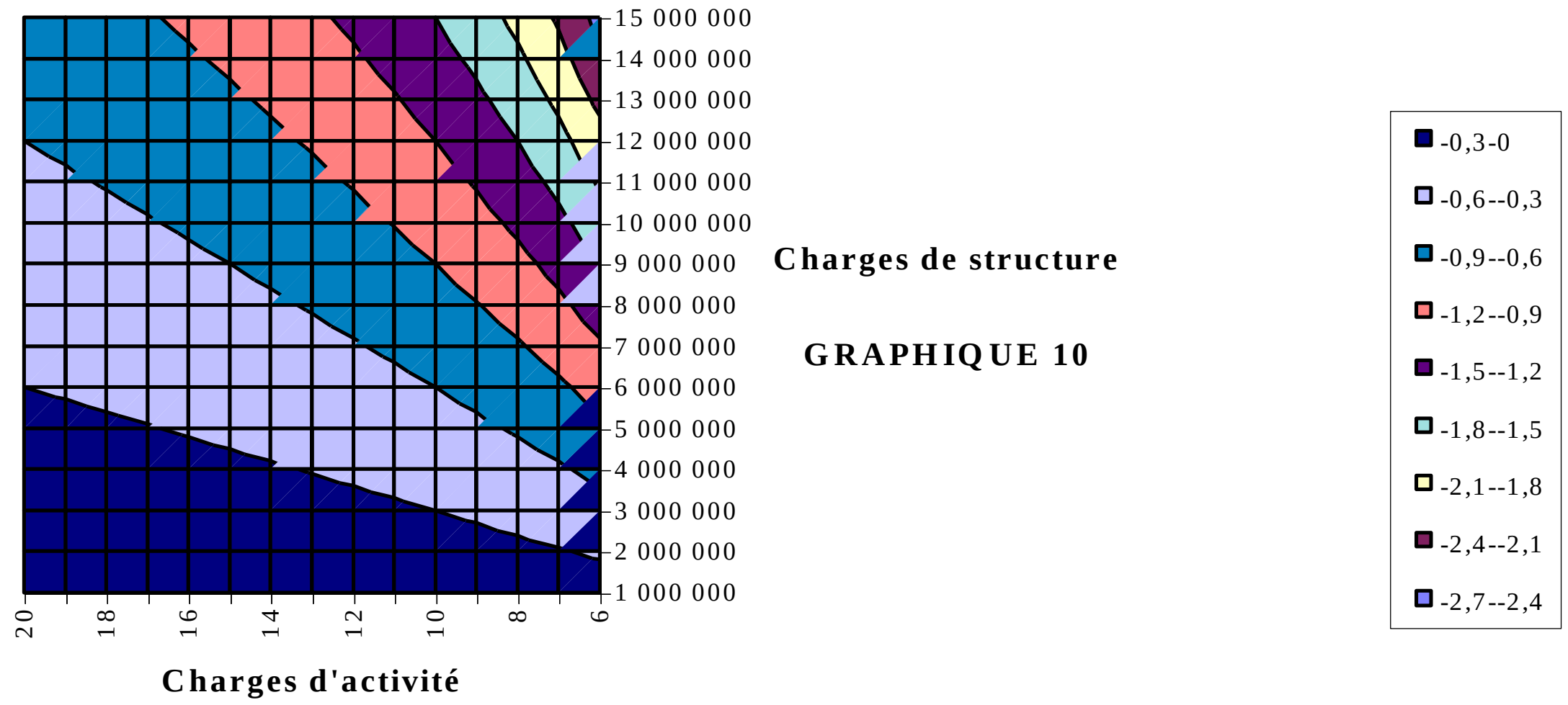

**GRAPHIQUE 10**

Ainsi pour une structure se caractérisant par f = 8 000 0000, v = 12, et en supposant b = 20 (prix du marché, constituant la valeur plafond de « v »), le coefficient d'évolution « a » égale - $1\times10^{-6}$. Par conséquent, si « f » doit augmenter à 10 000 000, « v » doit diminuer au moins à 10. L'élasticité est de - 0,66, comme on peut s'en assurer sur les graphiques 9 et 10. Dans ce cas, la baisse requise des coûts variables est limitée : si « f » augmente de 1%, « v » doit diminuer de 0,66%. Mais si l'on se situe à un niveau de coûts fixes supérieur, et/ou à un niveau de coûts variables inférieur l'équilibre de la trésorerie sera plus difficile à conserver car la valeur de l'élasticité sera plus forte.

Remarquons, en outre, que pour apprécier le risque d'illiquidité sous son aspect immédiat et sous son aspect à terme, une analyse de l'évolution des coûts fixes décaissables par rapport aux charges calculées doit compléter l'étude. Par exemple, l'effet de levier de trésorerie immédiate se détériorera si l'augmentation des coûts fixes décaissables est plus rapide que celle coûts fixes totaux.

Chaque niveau de coûts fixes et chaque niveau de coûts variables associé se caractérisent donc par un coefficient d'élasticité. Cette approche de l'élasticité de « v/f » peut être qualifiée de « relative », parce que caractéristique du type de combinaison productive. Cependant, le graphique 11 suivant met en évidence que quelle que soit l'ampleur de la variation des coûts à partir d'un couple (f, v) donné le coefficient d'élasticité est constant, à condition bien entendu que la baisse des coûts variables par rapport à l'augmentation des coûts fixes réponde bien aux paramètres de la relation v = af + b. Cette seconde approche de l'élasticité peut être qualifiée d'« absolue ».

Le cas particulier présenté sur le graphique 11 montre que l'élasticité « relative » calculée pour chaque valeur du couple ( f, v ) diminue de -0,05 à -2,5, alors que l'élasticité « absolue », c'est à dire celle calculée pour toute ampleur de variation des coûts fixes à partir d'un même niveau de coûts fixes et pour un même couple de départ ( f, v ) - ici à partir de la base f = 2 000 000 et v = 20 - engendre une élasticité constante, « - 0,05 » en l'occurrence. En d'autres termes, pour chaque couple ( f, v ) correspond une valeur d'élasticité relative à partir de laquelle existe une élasticité absolue de « v » par rapport à « f », de même valeur. La transformation de la structure des coûts suite à un changement de combinaison productive doit suivre un chemin critique, l'élasticité absolue, pour éviter une détérioration de la sensibilité de la trésorerie. Mais du même coup cette transformation modifie l'élasticité relative pour donner à la nouvelle combinaison une sensibilité plus forte.

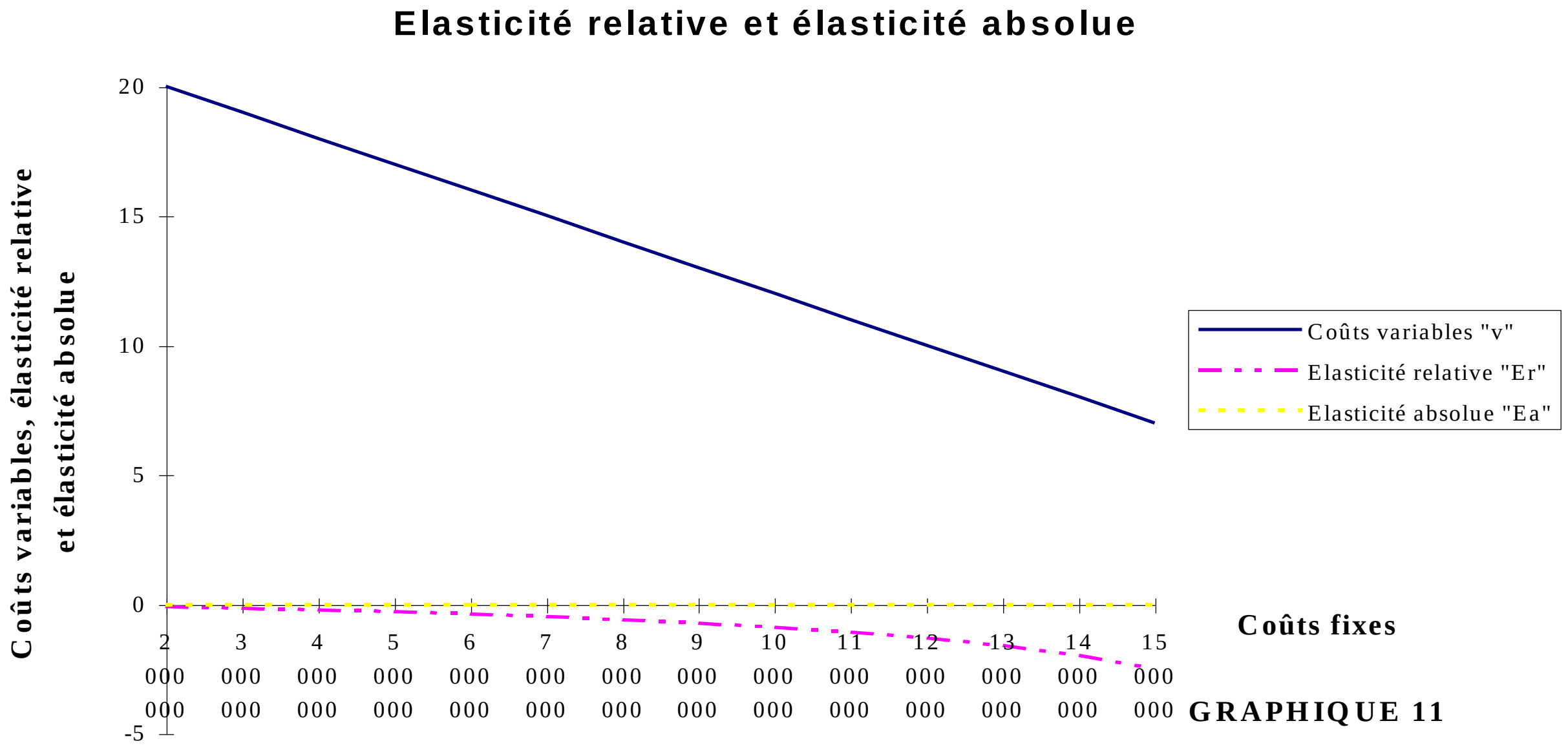

GRAPHIQUE 11

La généralisation à toutes les variations possibles de « v » par rapport à toutes les variations possibles de « f » pour un couple ( f, v ) donné correspond à la représentation graphique n° 12 suivante :

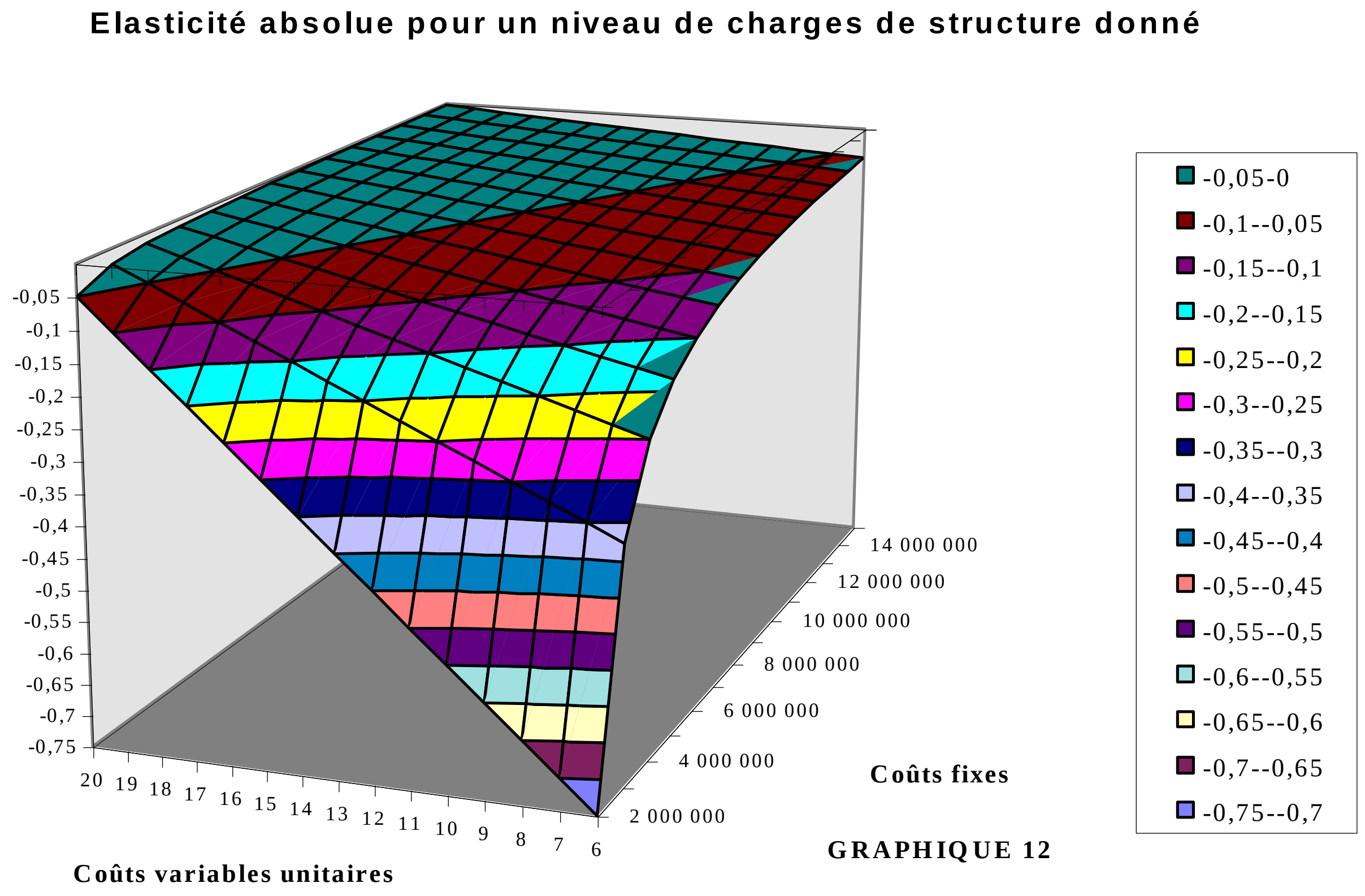

GRAPHIQUE 12

Une vue « plan » - graphique 13 - donne une meilleure vision de la succession de droites d'indifférence, élasticité absolue, associant à une même valeur du coefficient d'élasticité relative caractéristique d'une combinaison productive initiale donnée un ensemble de couples de variation « coûts fixes-coûts variables », (Δv, Δf), équivalents du point de vue de la liquidité de l'entreprise.

**Elasticité absolue ( projection )**

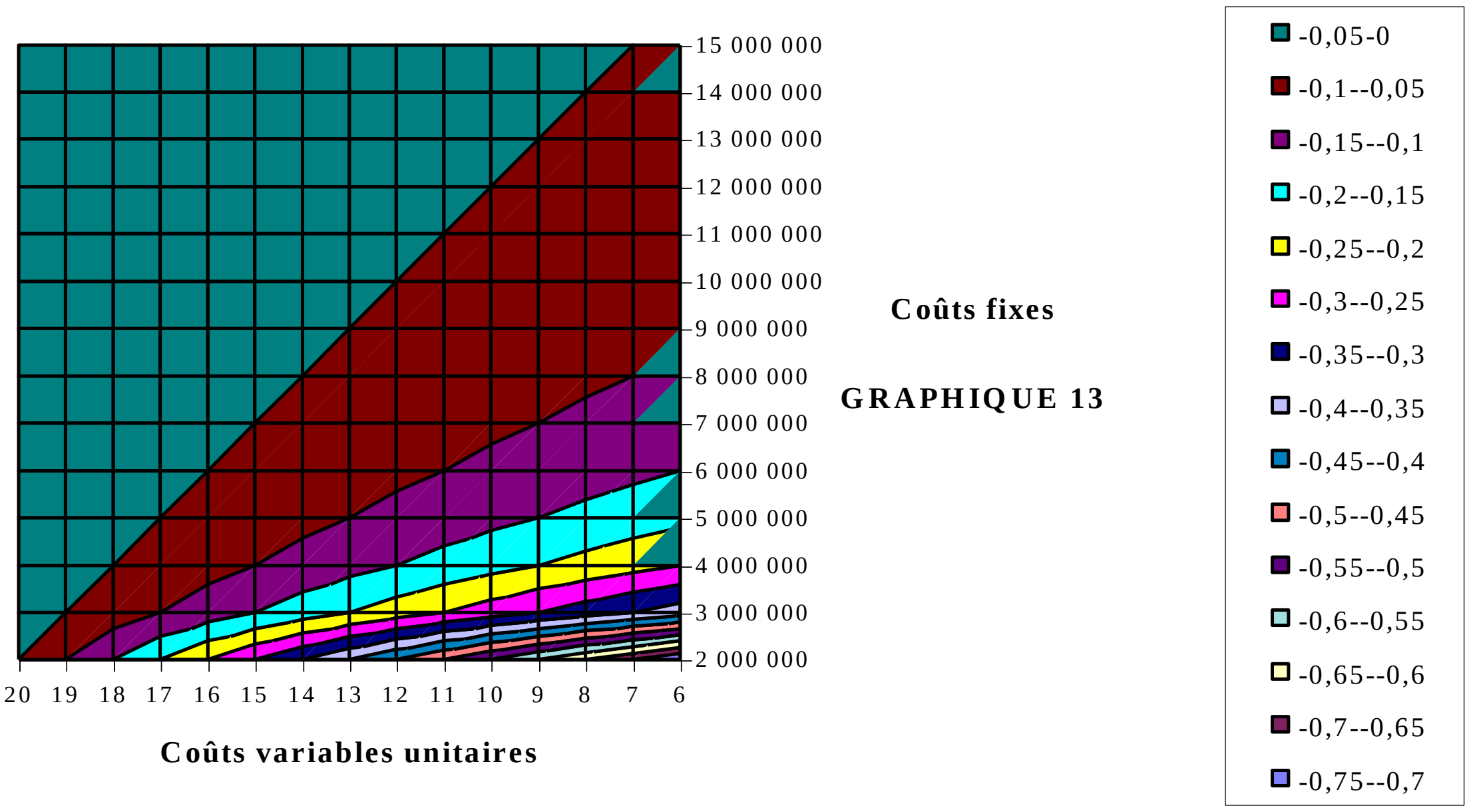

**GRAPHIQUE 13**

Nous remarquerons que plus les taux de croissance de «f » sont élevés plus les coefficients d'élasticité absolue se rapprochent de zéro quelle que soit la variation résultante de « v ». Cette représentation d'une situation particulière est à rapprocher du tableau des Plages d'élasticité « Δv/v, Δf/f » (page 16) qui rend compte du phénomène dans sa globalité. Le taux de décroissance des charges d'activité étant limité à - 1, un faible taux de croissance de « f » engendre une élasticité qui tend vers [-1, -∞], un fort taux de croissance donne une élasticité qui tend vers [-1, 0]. Dans notre exemple, compte tenu d'une élasticité absolue de 0,05, l'équilibre ne semble pas trop difficile à maintenir. Par contre si la réaction de « v » à une variation de « f » devait évoluer, et ne respecter plus le comportement des paramètres prévus dans v = af + b, alors l'élasticité absolue changerait de ligne d'indifférence pour se situer à un niveau supérieur, moins favorable, ou inférieur, plus favorable. Il pourrait donc alors devenir difficile à l'entreprise de conserver son degré de liquidité antérieur. Les situations les plus risquées sont celles où une augmentation de faible ampleur de « f » doit être compensée par une forte baisse de « v ». Ainsi si Δv/v doit tendre vers - 1 pour Δf/f tendant vers zéro l'élasticité est très forte et donc l'équilibre est plus difficile à maintenir. Il n'est pas certain que les coûts variables puissent dans toutes les situations être réduits dans une aussi large proportion.

En résumé, pour chaque type de combinaison productive caractérisé par un couple ( f, v ), existe une élasticité spécifique - dite « relative » - qui indique le degré de sensibilité de la structure (voir graphiques 9 et 10). Plus la combinaison est « labour saving » plus son élasticité caractéristique est forte. En outre, pour chaque changement de combinaison productive le degré de sensibilité de la structure de production - dite « élasticité absolue » - n'est maintenu que si les variations du couple (f, v) suivent leur chemin d'indifférence (voir graphiques 12 et 13). Dans le cas contraire la sensibilité de la liquidité peut se détériorer si la variation des charges d'activité est moins forte que prévu, où s'améliorer si leur variation est plus importante.

Dans notre exemple les limites entre lesquelles peut évoluer l'élasticité « absolue » par rapport à sa valeur critique - celle qui maintient la sensibilité présente de la trésorerie associée à un couple (f, v) - sont :

| | |
|---|---|
| Limite de détérioration de la sensibilité | ~ 0 |
| **Elasticité absolue « critique »** | **- 0,05** |
| Limite d'amélioration de la sensibilité | - 0,70 |

Connaissant la valeur relative de l'élasticité des charges d'activité par rapport aux charges de structure caractéristique d'une combinaison productive, il devient possible d'évaluer le risque d'insolvabilité afférent. L'effet de levier de trésorerie dépend comme nous l'avons vu dans la première partie de l'évolution des rapports f/m et f/Q. Une méthode d'évaluation du risque d'insolvabilité doit donc à partir de la connaissance des contraintes d'évolution de la combinaison productive que nous venons d'analyser, définir le jeu consécutif de ces deux paramètres f/m et f/Q.

L'objectif de la méthode est d'indiquer à l'entrepreneur les conditions dans lesquelles il peut transformer son processus productif et développer son activité pour parvenir à des performances plus importantes sans courir le danger d'une crise de trésorerie.

**B - Méthode d'évaluation du risque d'insolvabilité attaché à la nature de la combinaison productive.**

L'approche du risque d'insolvabilité est différente si l'on considère un changement de combinaison productive dans le cadre d'une capacité de production donnée ou au contraire lors d'une augmentation de celle-ci.

**a) Le risque d'insolvabilité de la combinaison productive pour une capacité de production donnée.**

Dans le cadre d'une capacité de production donnée, on peut déterminer les conditions à respecter pour que la sensibilité de la trésorerie ne se détériore pas. En effet, toute augmentation des coûts fixes doit s'accompagner d'une diminution des coûts variables, ou plus précisément d'un accroissement de la marge unitaire sur charges d'activité, comme nous l'avons déjà remarqué. Pour qu'il n'y ait pas détérioration de la sensibilité de la trésorerie il faut que l'élasticité de la marge unitaire sur coûts variables par rapport aux coûts fixes, au seuil de liquidité (1), soit égale ou supérieure à 1.

Au seuil de liquidité, noté Q*, nous avons $Q^* = f / m \Rightarrow E^*_{m/f} = 1$, et, $\mathbf{E_{m/f} \geq 1}$

Si l'on suppose que le prix sur le marché est inchangé, p = Cte, seul le contrôle de la variation des coûts variables unitaires, v, peut permettre de maîtriser la sensibilité de la trésorerie. Plus précisément on peut calculer la valeur limite que doit prendre v pour une augmentation de f. Au seuil de liquidité Q*,

$$Q^*p = Q^*v + f, \text{ d'où :}$$

$$(p-v) = f/Q^*$$

$$p- f/Q^* = v$$

$$\mathbf{v = - (1/Q^*)f + p}$$

(1) Seuil de liquidité à terme ou immédiate selon que l'on tienne compte ou non des charges calculées.

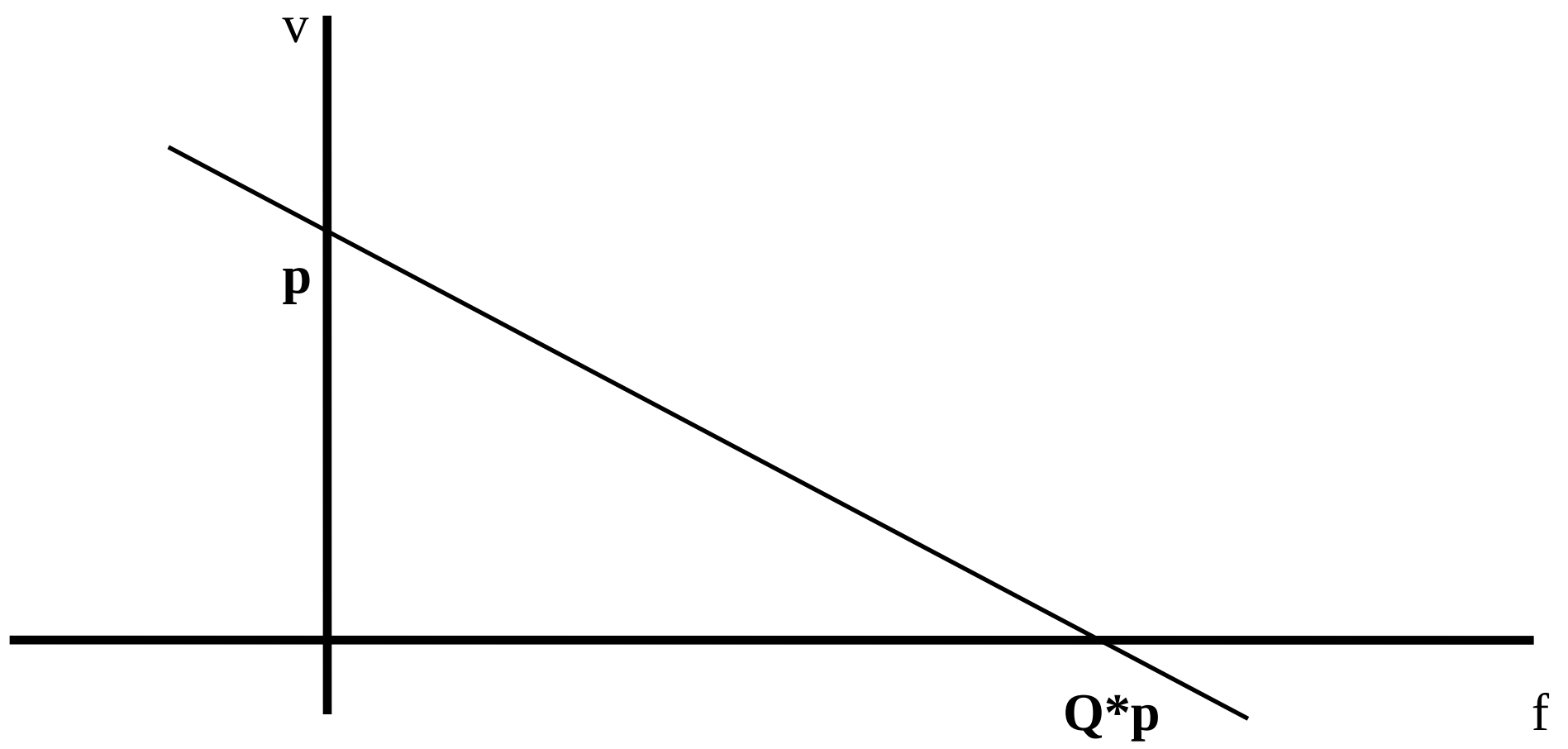

L'élasticité optimale E* des coûts variables « v » par rapport aux coûts fixes « f » s'écrit :

$$\mathbf{E}^*_{v/f} = \frac{f}{f - Q^*p}$$

où « f » est inférieur à Q*p, par définition. La représentation graphique est la suivante :

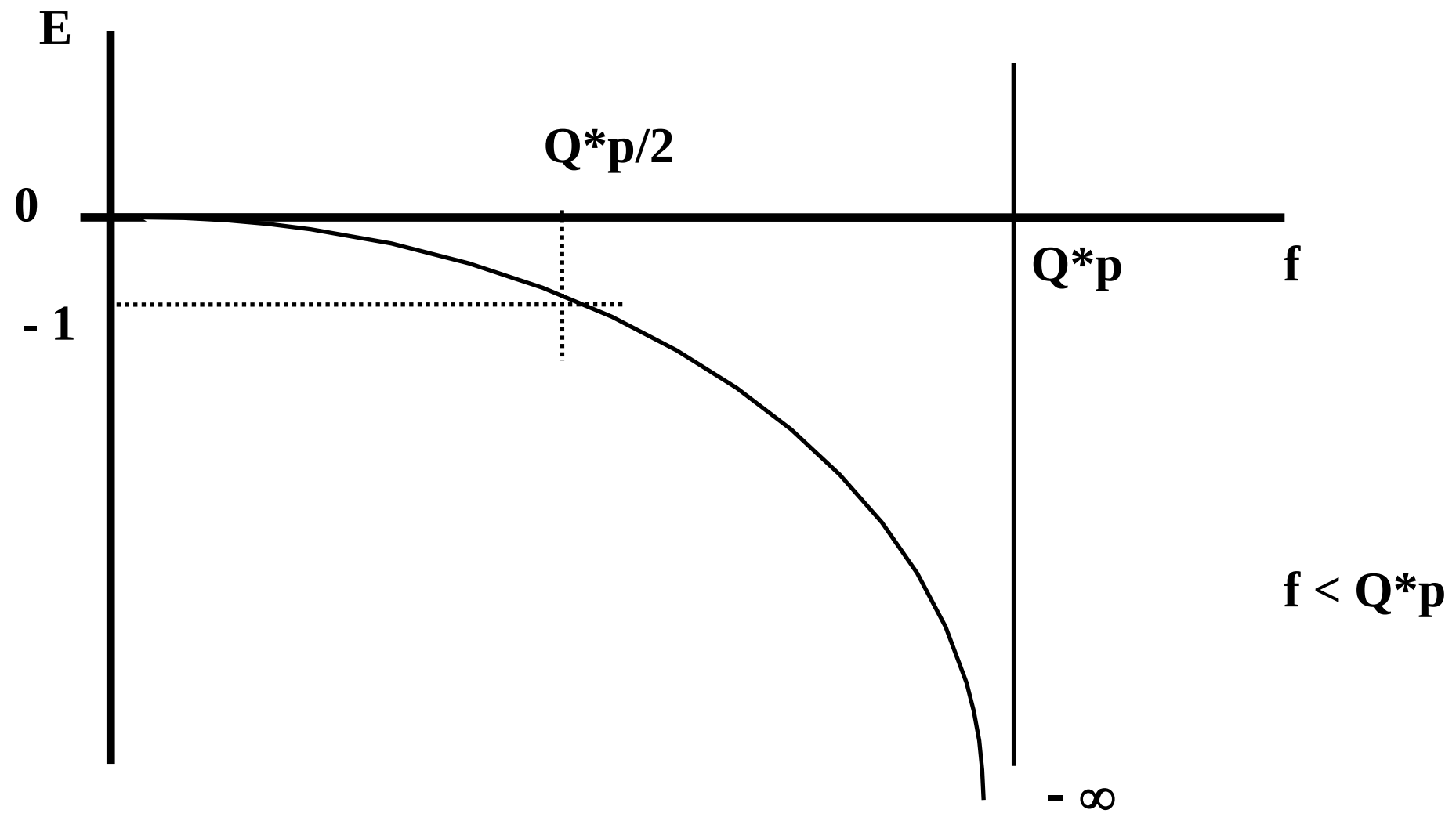

Ainsi, dans notre exemple l'entreprise ayant son seuil de liquidité immédiate égal à 250 000, l'élasticité optimale est :

$$E_{li}{}^*_{v/f} = \frac{2\,000\,000}{2\,000\,000 - (250\,000 \times 20)} = -0{,}666$$

Pour une augmentation des coûts fixes décaissables de 100 %, les coûts variables doivent au moins diminuer de 66,6 %, soit :

| $f_0$ **décaissables** | $v_0$ | $m_0$ | $\Delta f$ **décaissables** | $\Delta v$ | $f_1$ **décaissables** | $v_1$ | $m_1$ |
|---|---|---|---|---|---|---|---|
| 2 000 000 | 12 | 8 | 2 000 000 | 8 | 4 000 000 | 4 | 16 |

Dans ces conditions le changement de structure d'exploitation ne fragilisera pas la liquidité immédiate de la firme.

Si en même temps les charges calculées s'accroissent de 50 % les coûts fixes totaux montent alors à :

Charges fixes décaissables 2 000 000 + ( 2 000 000 ×100 % ) = 4 000 000
Charges calculées 6 000 000 + ( 6 000 000 × 50 % ) = 9 000 0000
Charges fixes totales = 13 000 000

Compte tenu de l'élasticité « optimale » au seuil de liquidité à terme, égale à :

$$E_{lt}*_{v/f} = \frac{8\ 000\ 000}{8\ 000\ 000 - (\ 1\ 000\ 000 \times 20\ )} = -\ 0,666$$

*( Les valeurs de $E_{li}$* et de $E_{lt}$* sont identiques puisque nous nous situons aux seuils de liquidité « optimaux » de la même structure d'exploitation )* et puisque l'augmentation des charges fixes totales est de 5/8 = 0,625, une diminution des coûts variables de : 0,625 × 0,666 = 0,41625, soit 5, s'impose :

| $f_{0\ totaux}$ | $v_0$ | $m_0$ | $\Delta f_{\ totaux}$ | $\Delta v$ | $f_{1\ totaux}$ | $v_1$ | $m_1$ |
|---|---|---|---|---|---|---|---|
| 8 000 000 | 12 | 8 | 5 000 000 | 5 | 13 000 000 | 7 | 13 |

Il apparaît donc que si la marge s'élève de 8 à 13, l'effet de levier de trésorerie à terme sera inchangé, alors que l'effet de levier de la trésorerie immédiate se détériorera. En revanche, si la marge sur coûts variables peut atteindre 16, alors la liquidité immédiate restera stable et la liquidité à terme s'améliorera, comme on peut s'en assurer dans les deux tableaux suivants :

**Evolution du seuil de liquidité immédiate**

| | $f_{0\ décaissables}$ = 2 000 000 | $f_{1\ décaissables}$ = 4 000 000 |
|---|---|---|
| $m_0$ = 8 | *250 000* | |
| $m_1$ = 13 | - | 307 692 |
| $m_1$ = 16 | - | **250 000** |

**Evolution du seuil de liquidité à terme**

| | $f_{0\ totaux}$ = 8 000 000 | $f_{1\ totaux}$ = 13 000 000 |
|---|---|---|
| $m_0$ = 8 | *1 000 000* | |
| $m_1$ = 13 | - | **1 000 000** |
| $m_1$ = 16 | - | 812 500 |

On peut préciser l'évaluation du risque d'insolvabilité en étudiant les conséquences de la variation des coûts variables sur la marge. Etant donné que le seuil critique de la liquidité s'exprime par « f / m », il n'y a pas de détérioration de la situation tant que « f », les coûts fixes, n'augmente pas plus vite que « m » la marge, pour un prix de vente unitaire, « p », et une quantité produite, « Q », inchangés.

Puisque m = p - v, on peut admettre que « m » est une fonction linéaire de « v », que l'on notera m = -v + p. La représentation graphique qui suit de cette fonction permet d'étudier le comportement de « m » par rapport à « v ».

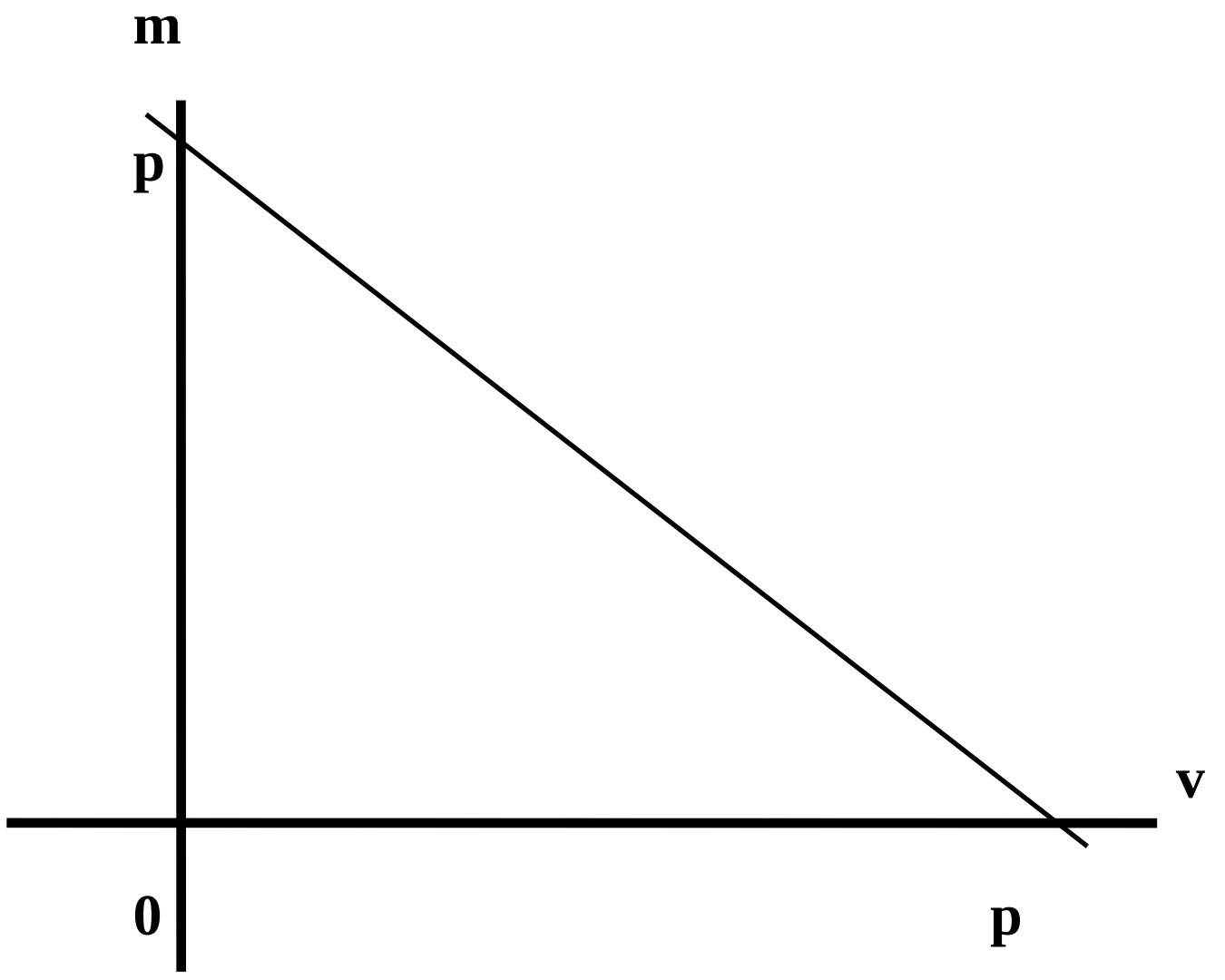

L'élasticité de « m » par rapport à « v » s'écrit :

$$\mathbf{E_{m/v} = -1 \; \frac{v}{-v + p} < 0}$$

et, s'analyse aisément à l'aide du graphique suivant :

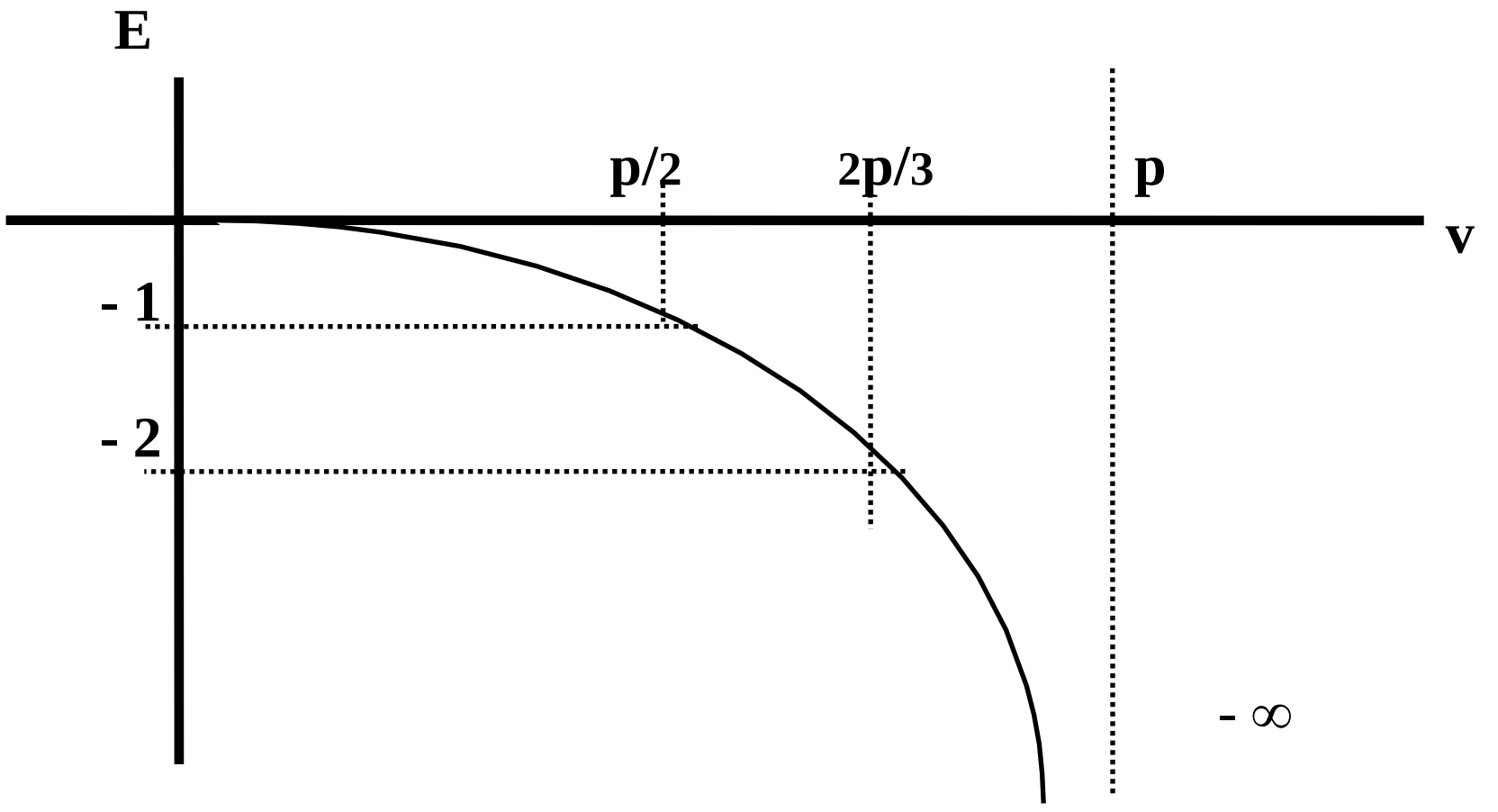

Par exemple, en reprenant les données précédentes nous pouvons calculer que pour un volume de production et un prix de vente, p = 20, fixés, un coefficient d'élasticité de la marge par rapport aux charges d'activité égale à :

$$E_{m/v} = \frac{-12}{-12 + 20} = 1,5$$

Ce qui permet de dire qu'une baisse de 1 % des coûts variables entraîne une augmentation de 1,5 % de la marge. Or, nous avons calculé que l'augmentation critique des coûts variables suite à un accroissement des coûts fixes a une élasticité de - 0,666. Par conséquent, si les coûts fixes augmentent de 100 %, les coûts d'activité doivent diminuer de 66,6 %, pour conserver à la liquidité son niveau antérieur, et la marge augmentera du même coup de 66,6 % × 1,5 ≈ 100 %. Dans notre exemple, des coûts fixes passant de 8 000 000 à 16 000 000 devront être compensés par une augmentation de la marge de 8 à 16, sans amélioration de la liquidité de la firme.

La croissance interne de l'entreprise donne à l'entrepreneur, si le marché est porteur, d'autres moyens de maîtriser le risque d'insolvabilité.

**b) Le risque d'insolvabilité de la combinaison productive lors de l'augmentation de la capacité de production.**

Lorsqu'il y a augmentation de capacité de production, on constate une augmentation des charges fixes. Pour que la sensibilité de la trésorerie ne s'aggrave pas il faut que les coûts fixes n'augmentent pas au-delà d'un certain plafond. Le graphique suivant représente de façon traditionnelle, les valeurs limites que peut prendre le coefficient d'élasticité des coûts fixes par rapport à la production :

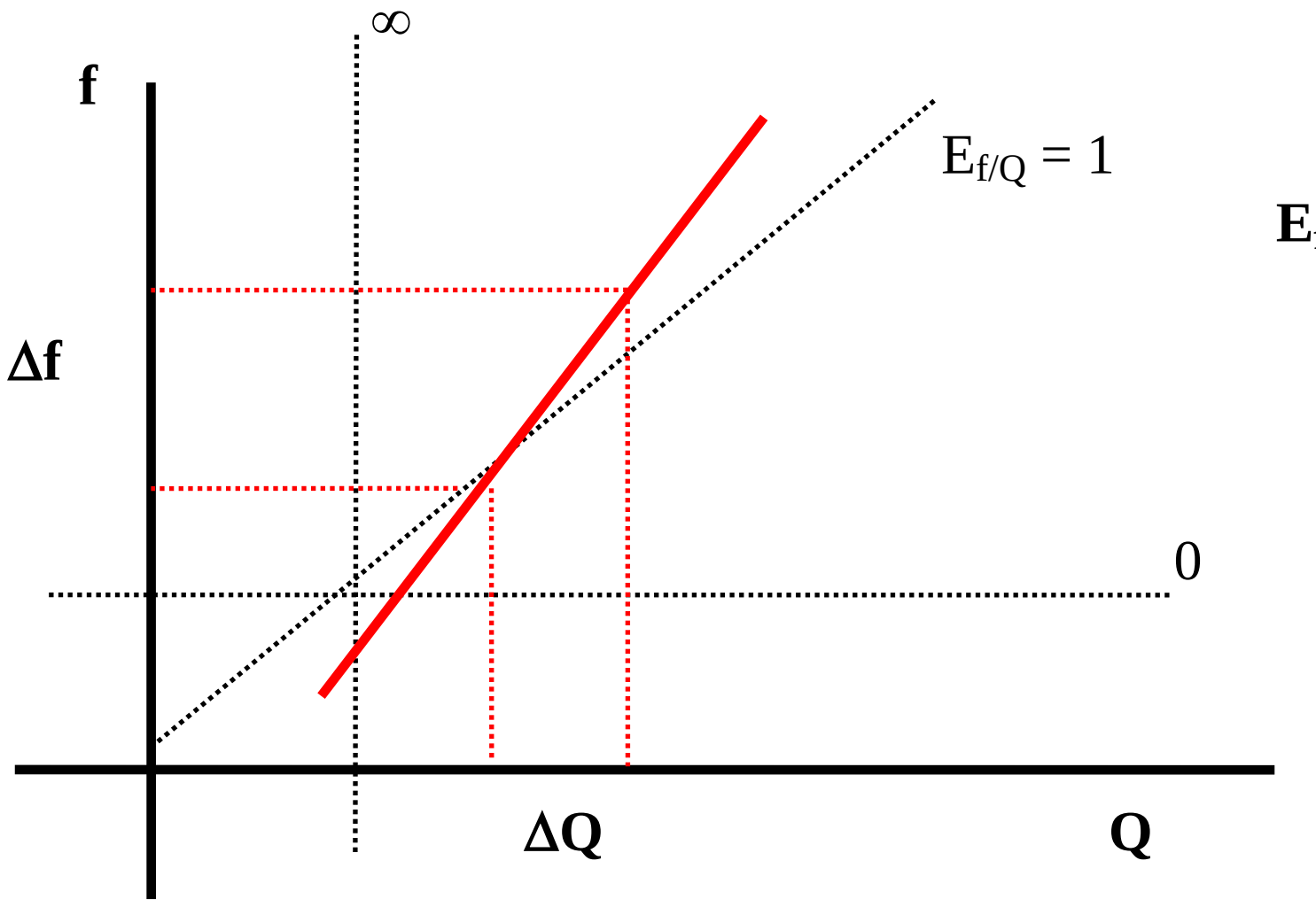

$$E_{f/Q} = \frac{\Delta f / f}{\Delta Q / Q}$$

Mais pour chaque valeur de Q le coefficient de levier varie. On doit donc rechercher l'élasticité « critique » ou « normative », noté $\mathbf{E_{f/Q}}$, qui permettra de définir ex-ante, la sensibilité recherchée de la trésorerie.

Si on note le chiffre d'affaires « Qp », les coûts variables « Qv », « f » les coûts fixes, et le résultat ou la CAF (1) « r », on peut écrire que :

$$\mathbf{Qp = Qv + f + r}$$

Or, la marge unitaire sur coûts variables « m », est égale à : m = p - v, par conséquent :

$$\mathbf{f = Qm - r}$$

que l'on peut représenter graphiquement :

(1) Selon que l'on retienne ou non les charges calculées.

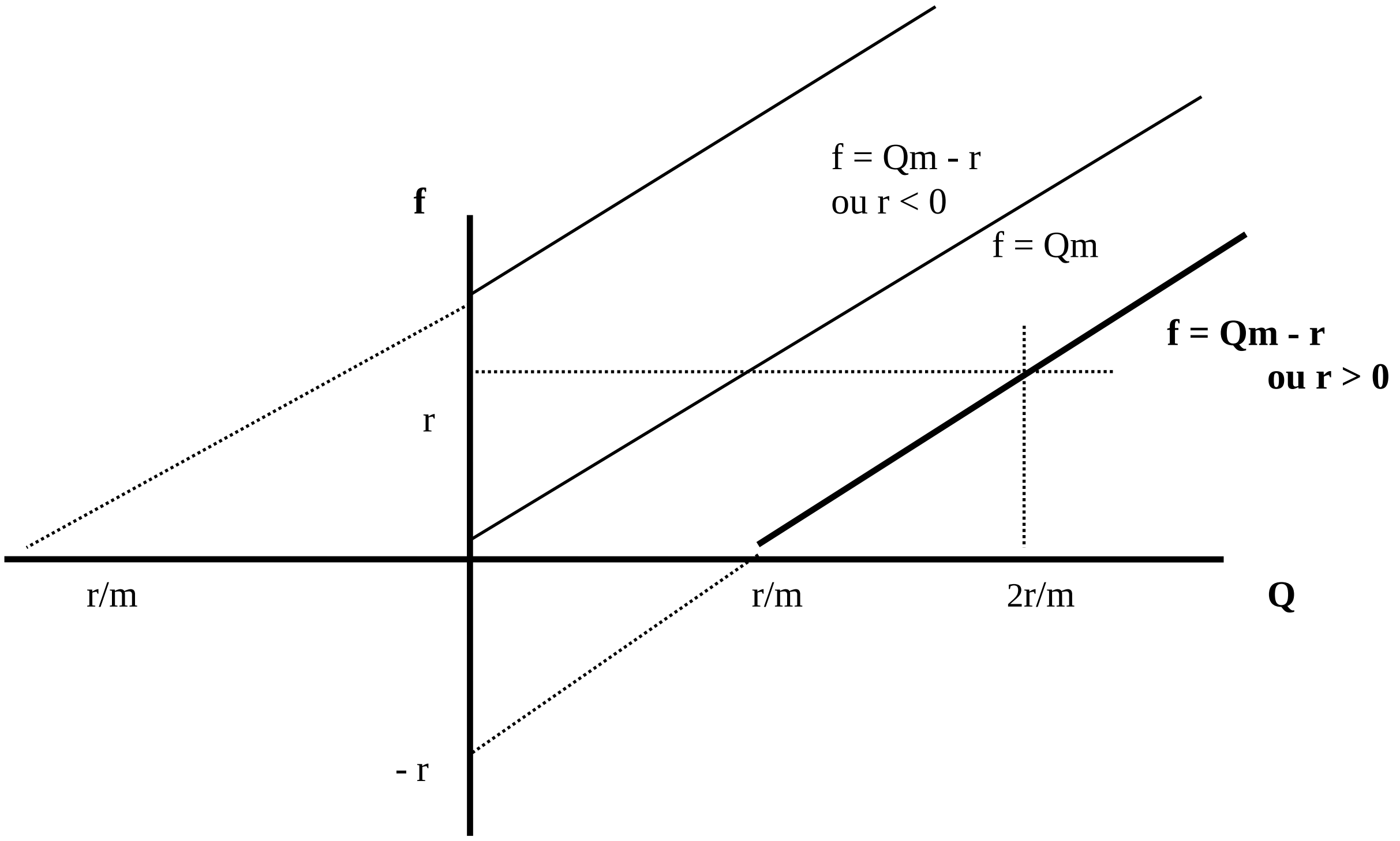

Si r = 0 $\Rightarrow$ $E_{f/Q} = 1$ $\forall$ $m \neq 0$

Si r > 0 , l'élasticité de « f » par rapport à « Q » s'écrit :

$$\mathbf{E_{f/Q} = \frac{Q}{Q - r/m}}$$

que l'on représentera graphiquement par :

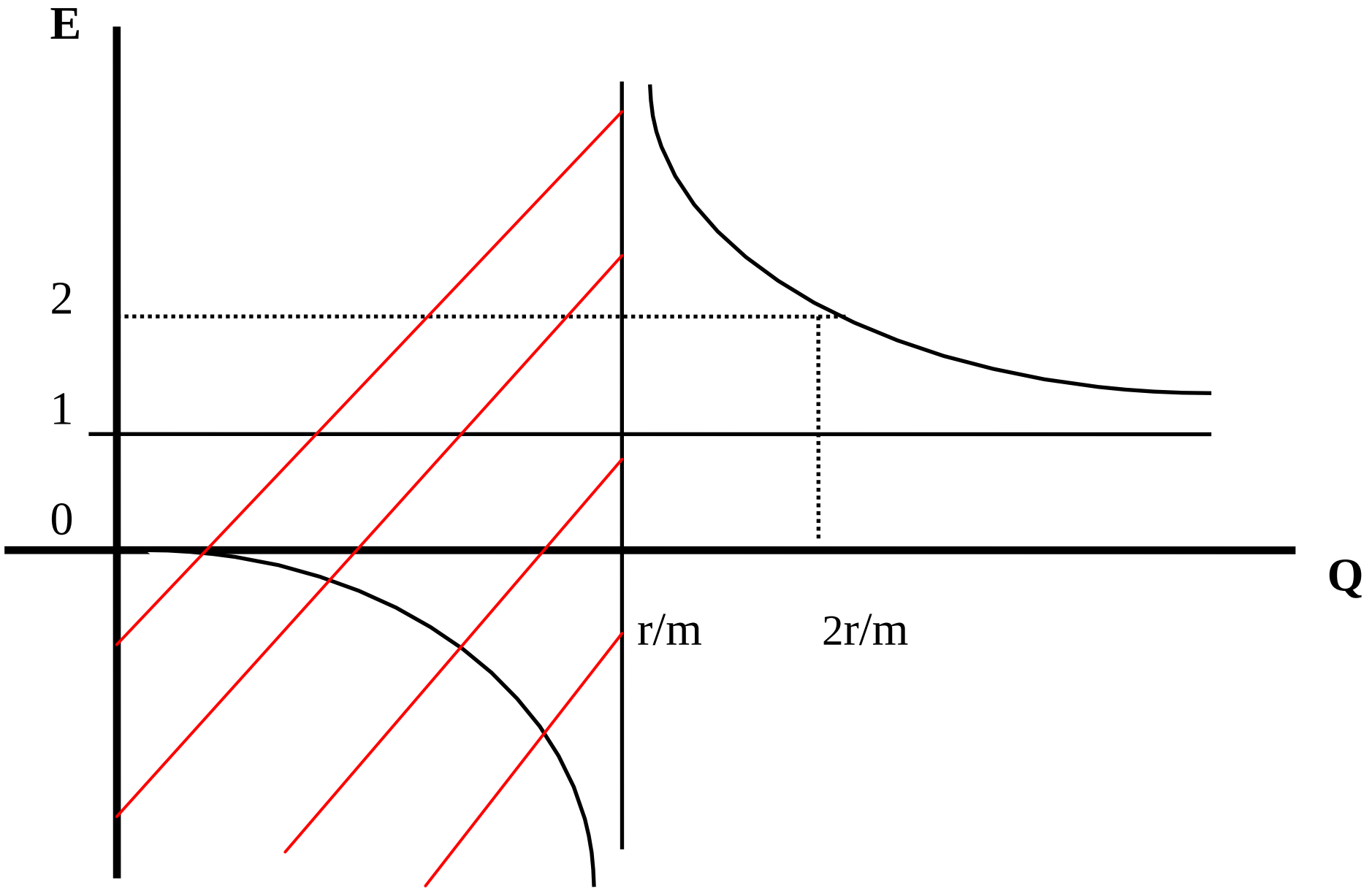

Seule l'élasticité positive à un sens économique puisque « f » varie dans le même sens que « Q », la capacité de production. Autrement dit, le dénominateur « Q - r/m » doit être positif, et par conséquent : Q > r/m , ou Qm > r ce qui signifie bien que la marge totale est supérieure au résultat.

Dans le cas où le résultat est une perte, r < 0, la variation de l’élasticité devient :

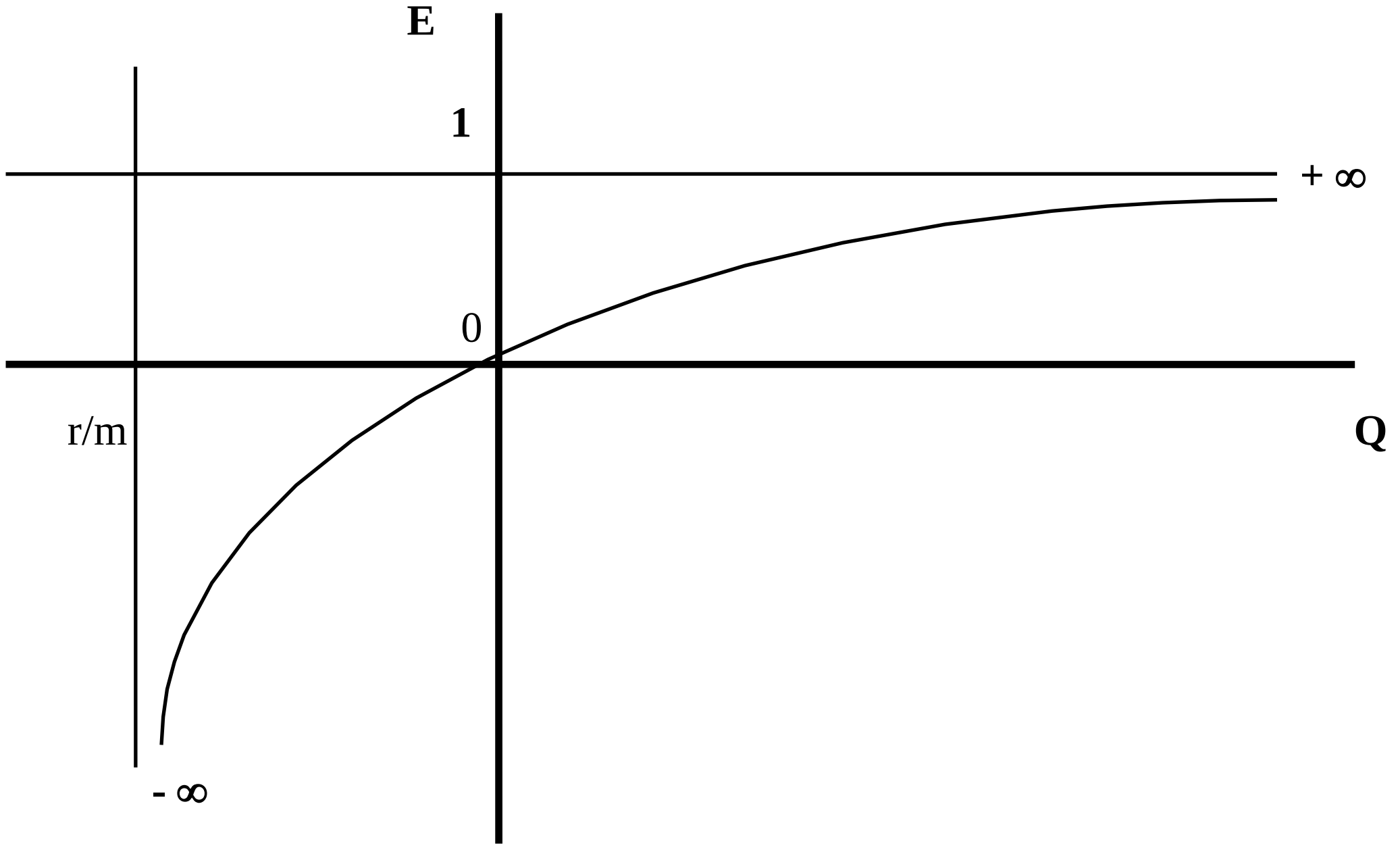

Comme on le voit, dans un contexte déficitaire l’élasticité de « f » par rapport à « Q » varie de 0 à 1.

Il convient de remarquer que le comportement de l’élasticité, $E_{f/Q}$, ci-dessus décrit ne vaut que pour un type de combinaisons productives donné : Qm = f + r. Lorsque l’entrepreneur change de combinaison productive la valeur de l’élasticité « f/Q » induite dépend du nouveau processus technologique retenu pour lequel tous les paramètres, f, Q, et v ont leur rapport modifié, et qui donnera une nouvelle relation Qm = f + r. Le schéma suivant représente les effets financiers d’une telle mutation technologique :

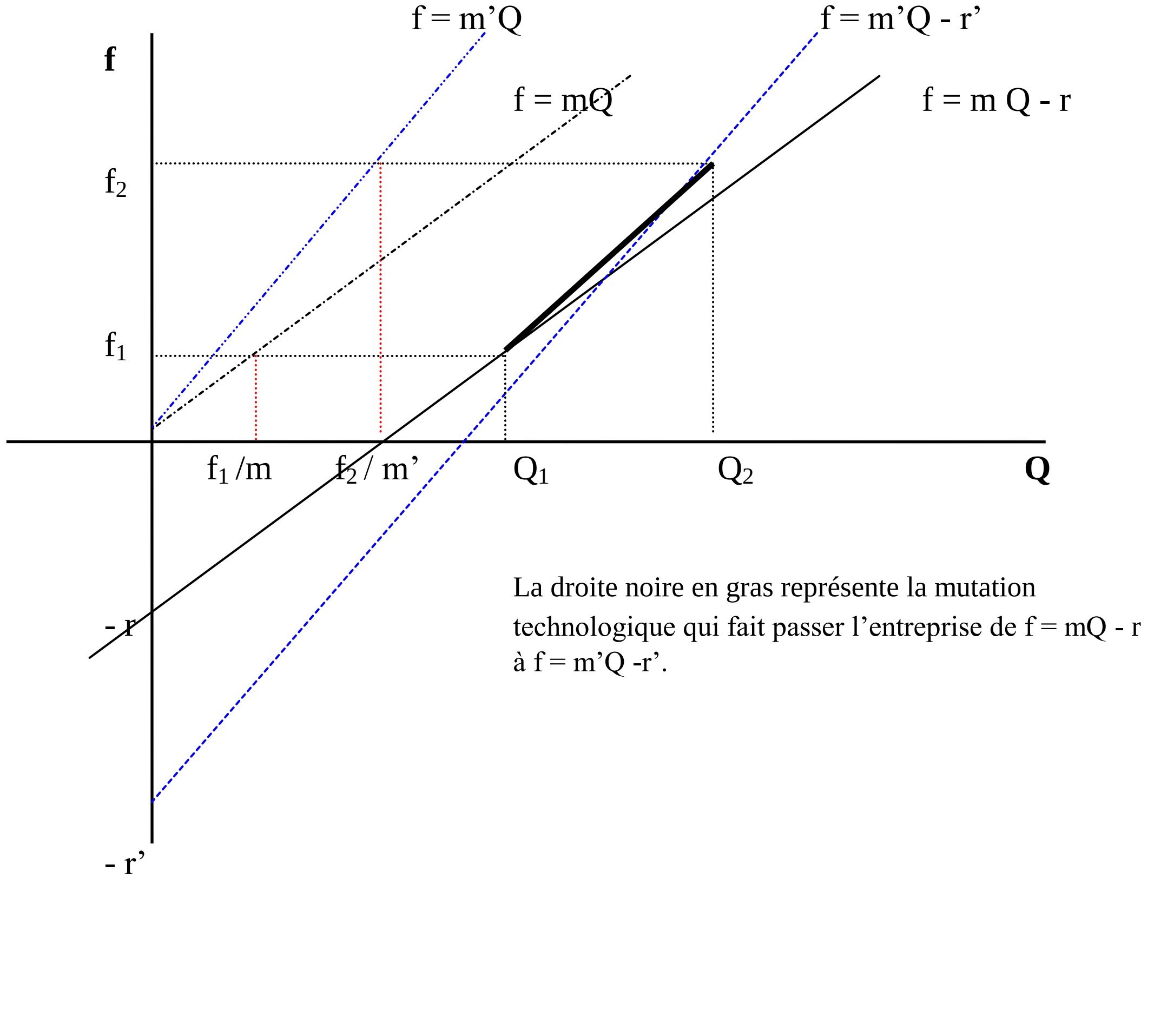

Le contrôle des paramètres se fera de la manière suivante, en fonction de la sensibilité de la trésorerie acceptée. Nous avons vu dans la première partie que :

$$E_{l/Q} = \frac{Q}{Q-(f/m)} \quad \text{et} \quad E_{l/m} = \frac{m}{m-(f/Q)}$$

A partir de ces deux relations on peut écrire que :

$$\mathbf{f = \frac{Qm\,(E - 1)}{E}}$$

Il convient de se rappeler que « f » peut désigner les coûts fixes totaux « $f_t$ » ou les coûts fixes décaissables « $f_d$ », que Q et m peuvent être des données ou des variables, et que « E » peut représenter quatre indicateurs :

- L'élasticité de trésorerie immédiate par rapport à la production (effet de levier d'exploitation),
- l'élasticité de trésorerie à terme par rapport à la production (effet de levier d'encaisse),
- l'élasticité de la trésorerie immédiate par rapport à la marge,
- l'élasticité de la trésorerie à terme par rapport à la marge.

Si l'augmentation de la capacité de production s'accompagne, en outre, d'une évolution favorable de la combinaison productive, la hausse de « f » entraîne une baisse de « v », dans les conditions étudiées précédemment. Alors, le prix de vente unitaire, « p », peut être baissé dans le cadre d'une politique de prix concurrentielle. L'entreprise sera plus compétitive sur son marché sans accroître le risque d'insolvabilité encouru, dans la limite de : p = m + v.

Illustrons ces considérations par un exemple en reprenant les données initiales. Une entreprise produisant 2 400 000 unités envisage un accroissement de 50 % de sa capacité de production. Cette politique implique une augmentation des charges calculées de 200 %, et de 20 % des charges fixes « décaissables ». Dans le même temps, les coûts variables unitaires baisseraient de 1/3. L'entreprise souhaite répercuter sur son prix de vente la baisse relative des coûts, sans aggraver la sensibilité de sa trésorerie.

La première analyse porte sur l'évaluation des effets de leviers de trésorerie acceptés, avant d'étudier les conséquences d'une éventuelle augmentation de la capacité de production accompagnée d'une transformation de la combinaison productive.

| **Paramètres de production** | **Valeur avant accroissement de la capacité de production** | **Valeur après accroissement de la capacité de production** |
|---|---|---|
| *Capacité de production* | 2 400 000 unités | 3 600 000 unités |
| *Charges calculées* | 6 000 000 F | 18 000 000 F |
| *Charges fixes décaissables* | 2 000 000 F | 2 400 000 F |
| *Charges fixes totales* | 8 000 000 F | 20 400 000 F |
| *Coûts variables unitaires* | 12 F | 8 F |
| *Prix de vente* | 20 F | 20 F |
| *Résultat* | 11 200 000 F | 22 800 000 F |
| *CAF* | 17 200 000 F | 40 800 000 F |

Les paramètres de l'effet de levier de trésorerie sont les suivant :

| **Indicateurs de la sensibilité de la trésorerie** | **Valeur avant accroissement de la capacité de production** | **Valeur après accroissement de la capacité de production** |
|---|---|---|
| *Seuil de liquidité immédiate* | 250 000 | 200 000 |
| *Seuil de liquidité à terme* | 1 000 000 | 1 700 000 |
| **Effet de levier d'encaisse** | 1,116 | 1,058 |
| **Effet de levier d'exploitation** | 1,714 | 1,894 |

Dans ce cas de figure on constate que la transformation de la combinaison productive qui accompagne l'augmentation de la capacité de production se traduit par une amélioration de la sensibilité de la trésorerie immédiate et une détérioration de la sensibilité de la trésorerie à terme. En appliquant la relation précédente on constate que le maintien de la liquidité à terme exigerait une augmentation du prix de vente à 21,60 F, alors que la trésorerie immédiate pourrait supporter une baisse du prix unitaire à 14,41 F. L'entrepreneur gérera cette opposition. Le graphique suivant permet d'envisager toutes les simulations possibles afin d'adapter l'évolution technologique aux contraintes du marché.

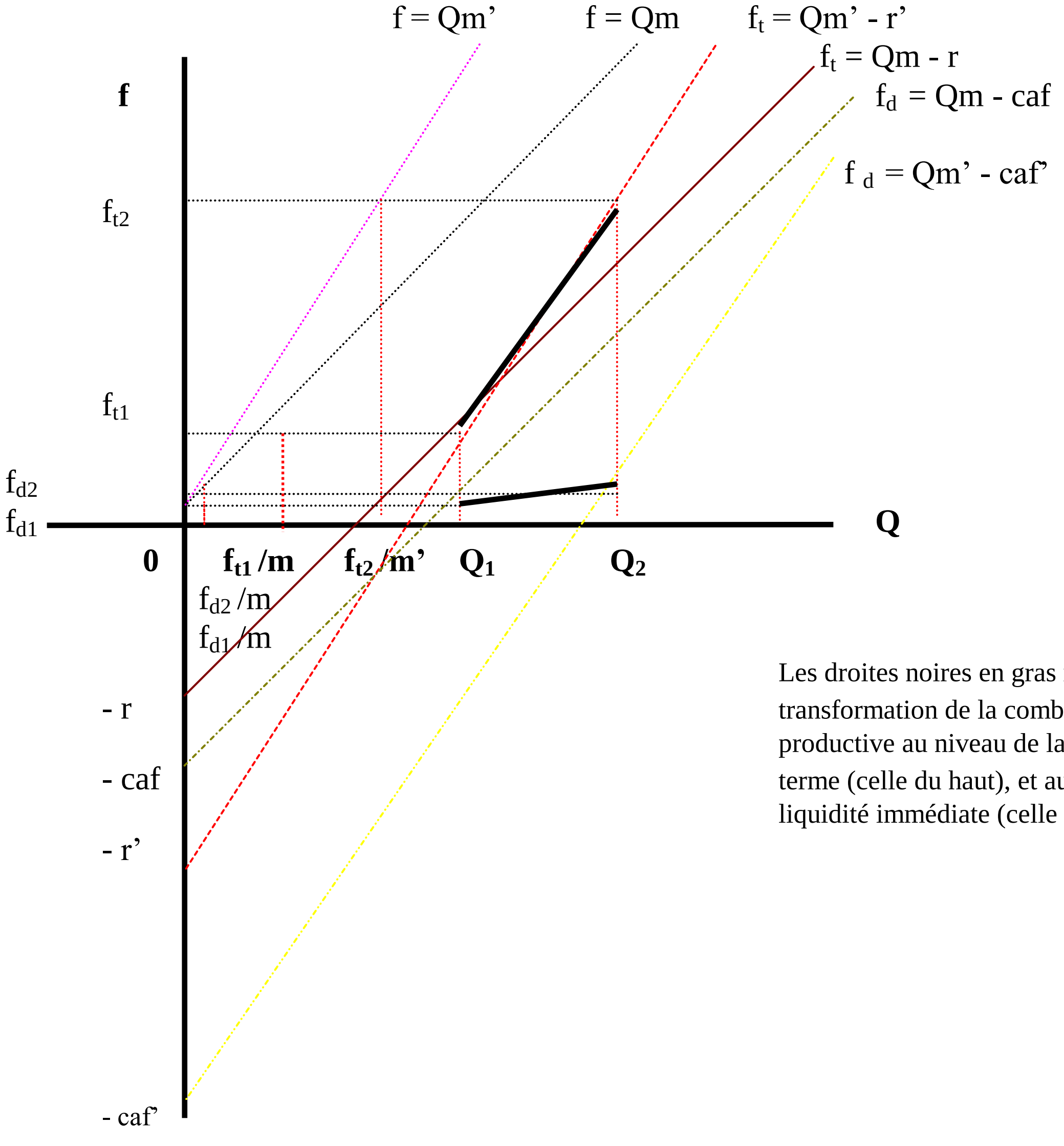

Les droites noires en gras représentent la transformation de la combinaison productive au niveau de la liquidité à terme (celle du haut), et au niveau de la liquidité immédiate (celle du bas).

Les données numériques, relevées dans les deux tableaux précédents, sont les suivantes :
$f_t = Qm - r$ : $f_t = 8Q - 11\,200\,000$ ( et, $f = Qm$ : $f = 8Q$ )
$f_t = Qm' - r'$ : $f_t = 12Q - 22\,800\,000$ ( et, $f = Qm'$ : $f = 12Q$ )
$f_d = Qm - caf$ : $f_d = 12Q - 17\,200\,000$
$f_d = Qm - caf'$ : $f_d = 12Q - 40\,800\,000$
$f_{t1} = 8\,000\,000$ ; $f_{t2} = 20\,400\,000$ ; $f_{d1} = 2\,000\,000$ ; $f_{d2} = 2\,400\,000$
$f_{t1} / m = 1\,000\,000$ ; $f_{t2} / m' = 1\,700\,000$ ; $f_{d1} / m = 250\,000$ ; $f_{d2} / m' = 200\,000$
$Q_1 = 2\,400\,000$ ; $Q_2 = 3\,600\,000$

La sensibilité de la trésorerie à terme (coefficient de levier d'exploitation) augmente de 1,714 à 1, 894. Ceci s'explique par le fait que sous l'effet de l'élasticité de « f / Q », le seuil de liquidité à terme s'est rapproché de la production totale. On peut le contrôler en comparant les relations permettant de calculer l'élasticité de la trésorerie à terme. Si la sensibilité de la trésorerie est inchangée :

$$\frac{Q_1}{Q_1 - (f_{t1}/m)} = \frac{Q_2}{Q_2 - (f_{t2}/m')}$$

Soit,

$$\frac{\mathbf{Q_1}}{\mathbf{Q_2}} = \frac{\mathbf{f_{t1} / m}}{\mathbf{f_{t2} / m'}}$$

On démontre que si le premier terme est inférieur au second, la sensibilité de la trésorerie s'améliore, et qu'elle se détériore dans le cas inverse ; en l'occurrence :

$$( 2\,400\,000 / 3\,600\,000 ) > ( 1\,000\,000 / 1\,700\,000 )$$
$$0{,}666 > 0{,}58$$

En ce qui concerne la trésorerie immédiate, on constate une diminution de son coefficient d'élasticité de 1, 116 à 1,058 que l'on pourrait expliquer et confirmer de façon analogue.

**Conclusion :**

L'association « capital-travail » détermine la structure des coûts de production. Compte tenu des contraintes du marché, offre et demande, la combinaison productive conditionne par conséquent la sensibilité de la trésorerie de l'entreprise.

Le tableau suivant rassemble les indicateurs de rupture de la liquidité d'une structure d'exploitation :

| | **Production** | **Marge** |
|---|---|---|
| **Coûts fixes décaissables** | Seuil de liquidité immédiate | Marge critique de liquidité immédiate |
| **Coûts fixes totaux** | Seuil de liquidité à terme | Marge critique de liquidité à terme |

Ils ont une signification spécifique dans la mesure où, d'une part, ils prennent comme référence soit la production, soit la marge, et d'autre part, ils retiennent soit la totalité des coûts fixes soit les coûts fixes décaissables seulement. A partir de ces indicateurs on définit un outil complexe d'analyse du degré de liquidité d'une combinaison productive : l'effet de levier de trésorerie. L'élasticité de la trésorerie immédiate mesure la sensibilité de la solvabilité « virtuelle » de la firme. L'élasticité de la liquidité à terme mesure la sensibilité de la trésorerie « virtuelle » de fin de période. En d'autres termes, la sensibilité des performances financières d'une combinaison productive dépend du niveau des coûts fixes, du volume de production et de la marge. Pour évaluer le risque d'insolvabilité attaché à une combinaison productive il faut connaître le comportement des charges opérationnelles vis à vis des charges de structure. La maîtrise des coûts, notamment la modulation des charges fixes, passe par le contrôle des facteurs qui génèrent ces coûts, « les inducteurs de coûts ». A cet égard la méthode « Activity-Based Costing » par la meilleure connaissance des causes de variation de consommation de ressources peut apporter une aide efficace à l'approche du risque d'insolvabilité par l'effet de levier de trésorerie. Il reste qu'au plan de la liquidité l'élasticité des coûts variables par rapport aux coûts fixes obère les mutations technologiques. Si la capacité de production demeure inchangée le contrôle de la variation de l'élasticité des coûts et de son incidence sur la marge unitaire est le seul paramètre à la disposition de l'entrepreneur pour maintenir la liquidité de la firme au niveau désiré. En revanche, si la capacité de production est une variable, l'entrepreneur peut utiliser le volume des ventes pour maîtriser sa trésorerie, mais alors la transformation de la combinaison productive doit être analysée afin de moduler dans la mesure du possible les éléments en jeu pour conserver à la structure de l'exploitation le degré de liquidité souhaité.

## Bibliographie :

BOULOT J.-L., CRETAL J.-P., JOLIVET J. et KOSKAS S., « Analyse et contrôle des coûts », Publi-Union, 1986.

COOPER R. et KAPLAN R. S., « Profit priorities from activity-Based Costing », Harvard Business Review, mai 1991.

DAROLLES Y., KLOPFER M., PIERRE F. et TURQ F., « La gestion financière », Publi-Union, 1986.

DEPALLENS G. et JOBARD J.-P., « Gestion financière de l'entreprise », Sirey, 10ème édition, 1990.

LAVOYER J.C., et TERNISIEN M., « Le tableau des flux de trésorerie », La Villeguérin Editions, Paris, 1989.

LEVASSEUR M. et QUINTART A., « Finance », 2ème édition, 1992.

LEVY A., « Management financier », Economica, 1993.

MARGERIN J. et AUSSET G., « Comptabilité analytique », Les Editions d'Organisation, 1993.

PATTISON D. D. et ARENDT C. G. : « Activity-Based Costing : It doesn't work all the time », Management accounting , avril 1994.

RICHARD J., « Les tableaux de flux du cycle », Banque, n° 456 décembre 1985.

TERNISIEN M., « L'importance du concept de flux de trésorerie disponible », R.F.C. n° 265 - mars 1995.

VAN HORN J., « Gestion et politique financière », tome 2, Dunod, 1985.